\documentclass[a4paper,11pt]{article}
\pdfoutput=1 
\usepackage{graphicx}
\usepackage{dcolumn}
\usepackage{bm}
\usepackage{amssymb}
\usepackage{graphicx} 
\usepackage{amsbsy}
\usepackage{amsmath} 
\usepackage{slashed}
\usepackage{amsfonts}
\usepackage{comment}
\usepackage{braket}
\usepackage{tensor}
\usepackage[scriptsize]{subfigure}
\usepackage{float}
\usepackage{enumerate}
\usepackage{multirow}
\usepackage{yfonts}
\usepackage{breqn}

\usepackage{jheppub}

\usepackage[utf8]{inputenc}

\usepackage{listings}
\usepackage{color}

\definecolor{mygreen}{rgb}{0,0.6,0}
\definecolor{mygray}{rgb}{0.5,0.5,0.5}
\definecolor{mymauve}{rgb}{0.58,0,0.82}

\definecolor{darkWhite}{rgb}{0.94,0.94,0.94}

\newcommand{\dd} {\mathrm{d}}
\newcommand{\tr} {\mathrm{tr}}
\newcommand{\Tr} {\mathrm{Tr}}

\newcommand{\al}{\alpha}
\newcommand{\be}{\beta}

\newcommand{\de}{\delta}
\newcommand{\De}{\Delta}

\newcommand{\eps}{\epsilon}

\newcommand{\sig}{ \sigma}

\newcommand{\Q}{\mathcal{Q}}
\newcommand{\iQ}{\mathcal{Q}'}
\newcommand{\GQ}{\mathcal{G}_\mathcal{Q}}
\newcommand{\toL}{\!\!\!\!}
\newcommand{\ct}{c}
\newcommand{\F}{{\mathcal{F}}}
\newcommand{\G}{{\mathcal{G}}}
\newcommand{\fb}{{\bar{f}}}

\usepackage{dsfont}

\usepackage{hyperref}

\usepackage{color}
\definecolor{verdes}{cmyk}{0.92,0,0.59,0.4}

\definecolor{Grn}{rgb}{0.1,0.5,0.2}
\definecolor{Blu}{rgb}{0.,0.,0.1}
\definecolor{Red}{rgb}{0.7,0.1,0.1}
\definecolor{SE}{rgb}{0.5,0,0.4}
\definecolor{Tur}{rgb}{0,0.75,0.65}

\newcommand{\Verdes}[1]{{\color{verdes}{#1}}}

\newcommand{\ds}[1]{\Verdes{[Diego:\,#1]}}

\usepackage{hyperref} 
\hypersetup{
     colorlinks = true,
     linkcolor = black,
     citecolor = blue,
     anchorcolor = blue,
     filecolor = blue,
     urlcolor = blue,
     bookmarks=true,
     linktocpage =true,
     }

\renewcommand{\vec}[1]{\mathbf{#1}}

\newcommand{\dime}{d}

\newcommand{\dx}[1][x]{{{\rm d} #1}}
\newcommand{\mathi}{{\text{i}}}

\newcommand{\colred}[1]{\textcolor{black}{#1}}

\usepackage{xparse}

\NewDocumentCommand{\dkpara}{O{k} O{\dime}}{\frac{{\rm d} {#1}^{\parallel}}{(2\pi)^{#2}}}
\NewDocumentCommand{\dkd}{O{k} O{\dime}}{ \frac{{\rm d}^{#2} {#1}}{(2\pi)^{#2}}}
\NewDocumentCommand{\dxd}{O{x} O{\dime}}{ {\rm d}^{#2} {#1} }
\NewDocumentCommand{\ad}{O{} O{}}{ \bigl \langle {#1}\bigr\rangle_{{\rm ad}{#2}} }

\begin{document}

\title{\boldmath Strong-field regime within effective field theory}

\author [1,2,3]{Sebasti\'an Franchino-Vi\~nas,}
\author [4]{J\'er\'emie Quevillon} 
\author [4,5]{and Diego Saviot}

\affiliation[1]{Departamento de Física, Facultad de Ciencias Exactas Universidad Nacional de La
Plata, C.C. 67 (1900), La Plata, Argentina}
\affiliation[2]{CONICET, Godoy Cruz 2290, 1425 Buenos Aires, Argentina}
\affiliation[3]{Universit\'e de Tours, Universit\'e d'Orl\'eans, CNRS, Institut Denis Poisson, UMR 7013, Tours, 37200, France
}

\affiliation[4]{Laboratoire d’Annecy-le-Vieux de Physique Th\'eorique,
CNRS – USMB, BP 110 Annecy-le-Vieux, 74941 Annecy, France}

\affiliation[5]{Laboratoire de Physique Subatomique et de Cosmologie, Universit\'e Grenoble-Alpes, CNRS/IN2P3, Grenoble INP, 38000 Grenoble, France}

\emailAdd{safranchino@fisica.unlp.edu.ar}
\emailAdd{jeremie.quevillon@lapth.cnrs.fr}
\emailAdd{diego.saviot@lapth.cnrs.fr}

\abstract{
 Building upon the Covariant Derivative Expansion, we develop a method to compute effective actions that is able to capture non-perturbative effects induced by strong background fields.   
We demonstrate the method in scalar QED, by deriving the full second-derivative corrections to the scalar Heisenberg--Euler effective action. The corresponding result is interpreted as an effective field theory with three characteristic scales, two of which are large (mass and field strength) in comparison with the remaining one (derivatives of the field). As an application, we show that, at this order, the transseries structure of the Schwinger pair production rate is preserved, even if the  involved coefficients are modified. Our analysis also helps clarify recent disagreements concerning the coefficients of this effective action. 

}

\maketitle

\section*{Conventions}
We will use the metric $\eta_{\mu\nu}=\operatorname{diag}(1,-1,-1,-1)$. We will work in natural units, for which $c=1=\hbar$.

\section{Introduction}

The quantum nature of the vacuum remains to a large extent \textit{terra incognita}, in particular when backgrounds fields are involved.
Focusing on quantum electrodynamics, a theoretical prediction is the Schwinger effect \cite{Schwinger:1951nm}, i.e. the production of an electron-positron pair by an electric field with order of magnitude $\frac{m_e^2c^3}{e\hbar}\sim 10^{18}\,\text{Vm}^{-1}$, which remains unattainable with current capabilities.
Still, technical advances in lasers allow us to reach unprecedented powers and have motivated the development of several experiments. To cite a few, the BIREF@HIBEF~\cite{Ahmadiniaz:2024xob} and the LUXE~\cite{LUXE:2023crk} proposals at DESY, the LASERIX collaboration in Paris~\cite{Kraych:2024wwd} and  groups at CoReLS in South Korea~\cite{Mirzaie:2024iey} have been conceived with the goal of observing the strong-field, quantum effects of the vacuum.

One can expect that the experimental developments will in the near future allow testing the region where non-perturbative effects like pair production become observable, and more precise theoretical predictions will be required.
In parallel, there currently exist scenarios within our reach where such effects could be observed. For instance, in Ref.~\cite{Schmitt:2022pkd} a mesoscopic variant of the Schwinger effect was realised and observed in graphene, while large (virtual) fields are involved in heavy-ion collisions~\cite{ATLAS:2017fur}. Moreover, it is believed that strong fields could have had a crucial rôle in early epochs of the universe, see \cite{Domcke:2021fee} and references therein.
Therefore, it can be argued that a better theoretical understanding of intense-field physics and the development of methods in this discipline are essential.

The pioneering works \cite{Heisenberg:1936nmg, Euler:1935zz} initiated the study of the strong-field regime by computing the effective action for a constant electromagnetic field which results upon integration of a fermionic field (see Ref.~\cite{Weisskopf:1936hya} for the scalar case).
Contrary to weak-field expansions---computed with methods such as diagrammatic matching, the Covariant Derivative Expansion (CDE)~\cite{Drozd:2015rsp, Henning:2014wua}, the perturbative heat kernel~\cite{Vassilevich:2003xt, Dunne:2007rt} and perturbative worldline techniques~\cite{Bastianelli:2024vkp}--- the Heisenberg--Euler effective action implies that the mass of the integrated-out field  is not necessarily much larger than the characteristic strength of the external fields. In particular, an imaginary contribution in the effective action naturally emerges in the non-perturbative computation, while to recover it from a perturbative expansion a resummation is required.

Currently, a variety of methods are being developed to study strong fields, including worldline resummed expressions~\cite{Bastianelli:2025khx, Fecit:2025kqb, Ilderton:2025umd}, worldline instantons~\cite{Semren:2025dix, DegliEsposti:2024upq}, 
the Furry expansion~\cite{Copinger:2024pai}, numerical quantum kinetic theory~\cite{Edwards:2025cco}, resurgence theory~\cite{Dunne:2022esi} and non-perturbative heat-kernel techniques~\cite{Franchino-Vinas:2025ejo, Franchino-Vinas:2023wea}. In parallel, some related expansions have been obtained, including the first terms in a derivative expansion of the effective action~\cite{Batalin:1971au, Lee:1989vh, Gusynin:1998bt} and the full-derivative approximation of effective actions for  weak fields~\cite{Boasso:2024ryt, Barvinsky:1990up}.

In this article, we will propose a novel alternative, based on an improvement of the CDE.  After the seminal papers by Gaillard~\cite{Gaillard:1985uh} and Cheyette~\cite{Cheyette:1987qz}, the method has lately shown its advantages when considering the effective field theory (EFT) approach to the Standard Model of particle physics~\cite{Henning:2014wua,Drozd:2015rsp}.
In a sense, the expansions that we are going to derive could be interpreted as an EFT with several scales, two of which could turn out to be of the same order for the operators of the theory. Indeed, the background field contributes with two scales, its strength and the size of its variations in spacetime, which have to be added to the scale given by the mass of the quantum field. In the strong-field regime, we are going to consider that both the field strength and the mass will be large, while derivatives of the background can be used to perform an expansion. This multiscale EFT approach reamins to a great extent an unexplored arena with potential  application in a variety of models.

Our method, which we will baptize the Strong-Field (SF) CDE, is naturally based on a momentum-space expansion. 
In Sec.~\ref{sect:expansion}, we will develop the formal features of this expansion in scalar QED, introducing a propagator which inherently contains the effects of the strong-field contribution of the electromagnetic field.

We will then apply the formalism to compute the corrections to the Heisenberg--Euler action arising from (small) inhomogeneities of the electromagnetic field. The results, which are contained in Sec.~\ref{sect:inhom}, show that the method is especially suited to perform a derivative expansion of the effective action when accompanied by symbolic algebra software. 
Furthermore, we analyse in detail the first order in this derivative expansion, including a discussion on the modifications induced on the Schwinger effect in Sec.~\ref{sec:Schwinger}. 
Some open questions are discussed in Sec.~\ref{sec:conclusions} and several useful intermediate computations are provided in the appendices.

\section{Modifying the CDE expansion}\label{sect:expansion}
\subsection{Gaillard--Cheyette CDE method}\label{CDE}

In the following, we are going to consider the theory of a massive charged (complex) scalar field $\phi$ coupled with an electromagnetic (classical) background $A_\mu$, living in a flat $d$-dimensional spacetime.
We also include a potential U involving combinations of light fields coupling quadratically to $\phi$.
The theory is defined by the action
\begin{align}
 S= \int \dxd[x][\dime] \,\phi^* \left(D^2+M^2+U\right)\phi\,,
\end{align}
where we have introduced the covariant derivative $D_\mu$ as customarily, 
\begin{align}
    {\colred D_\mu \equiv \partial_\mu + A_\mu}\,.
\end{align}

Our goal is to compute the effective action resulting from the integration of the quantum field, which can be defined by the functional integral
\begin{align}
e^{\mathi S_{\mathrm{eff}}}\equiv\int\mathcal{D}\phi\,e^{\mathi S}\,,
    \qquad
    S_{\mathrm{eff}}=\mathi \Tr \operatorname{Log}\left(D^2+M^2+U\right).
    \label{eq:TrLog}
\end{align}

Taking this as a starting point, we will recall the usual steps of the CDE. This will help to understand the improvements that we will introduce and  will grant access to the strong-field regime; further information on the perturbative approach can be found in Ref.~\cite{Drozd:2015rsp, Henning:2014wua}.
The first step in the evaluation of $S_{\mathrm{eff}}$ is to introduce  a plane-wave basis to evaluate the trace in momentum space.
Then, a key procedure for this work is the Gaillard--Cheyette trick based on the identity
\begin{align}\label{eq:gaillard}
    \int \mathrm{d}^dx\,\mathrm{d}^dq \,B
    =\int \mathrm{d}^dx\,\mathrm{d}^dq \,e^{-\mathi D_\mu\partial_q^\mu} B\, e^{\mathi D_\mu\partial_q^\mu} \,,
\end{align}
which is valid for a general $B=B(q,x-\mathi \partial_q)$, see Refs.~\cite{Gaillard:1985uh,Cheyette:1987qz}.
Evaluating the right-hand side of Eq.~\eqref{eq:gaillard} with the Baker--Campbell--Hausdorff formula, the result is
\begin{align}
 \mathcal{L}_{\rm eff}&=\mathi \int \frac{\dxd[q][\dime]}{(2\pi)^\dime} \tr \log \left[-(\tilde F_{\nu\mu} \partial_q^\mu+ q_\nu )^2+M^2 +\tilde U \right], \label{eq:lagrangian_GC}
\\
\label{eq:tilde_F}\tilde F_{\mu\nu}&\equiv \sum_{k=0}^{\infty} \frac{(k+1)\mathi^k}{(k+2)!} [D_{\alpha_1},[\cdots [ D_{\alpha_k},F_{\mu\nu}]\cdots] \partial_q^{\alpha_1\dots\alpha_k},
\\
\label{eq:U_tilde}\tilde U&\equiv \sum_{k=0}^{\infty} \frac{\mathi^k}{k!} [D_{\alpha_1},[\cdots [ D_{\alpha_k},U]\cdots] \partial_q^{\alpha_1\dots\alpha_k},
\end{align}
where $\partial_{q}^{\alpha_1\cdots \alpha_n}\equiv \partial_q^{\alpha_1}\dots\partial_q^{\alpha_n}$ and the field strength $F_{\mu\nu}$ is defined as
\begin{align}
    \colred{F_{\mu\nu}\equiv -[D_\mu, D_\nu]}\, .
\end{align}
Notably, the action of the background field is now encoded in a series of commutators that are manifestly gauge-invariant.
The expression for $\tilde F_{\nu\mu} \partial_q^\mu$, according to the definition in Eq.~\eqref{eq:tilde_F}, is equivalent to a momentum-space version of the gauge potential in the Fock--Schwinger gauge~\cite{Fock:1937, Pascual:1984zb}. Notice, still, that we have not imposed \emph{a priori} any gauge condition; instead, in the computation we have expressed all the commutators in a covariant way, expanding around a given point. 
Here we consider an abelian theory, for which all the commutators can be readily computed in closed form,
\begin{align}\label{eq:Ftilde}
\tilde F_{\mu\nu}&= \sum_{k=0}^{\infty} \frac{(k+1)\mathi^k}{(k+2)!}  \partial_{\alpha_1\dots\alpha_k} F_{\mu\nu} \partial_q^{\alpha_1\cdots \alpha_k}.
\end{align}

The usual perturbative way to proceed, described in Ref.~\cite{Drozd:2015rsp}, is to rewrite the logarithm as an integral of an appropriate propagator over the mass; expanding in powers of this propagator, $\Delta\equiv\frac{1}{q^2-m^2}$, we get a weak-field expansion:
\begin{align}\label{eq:CDE-wf}
    \mathcal{L}_{\rm eff}=-\mathi\int \frac{\dxd[q][\dime]}{(2\pi)^\dime} \int^{M^2} \dx[m^2]\sum_{n=0}^\infty \left[-\De \left(\{q_\mu,\tilde F_{\nu\mu}\}\partial_q^\nu+\tilde F_{\sig\mu}\tilde F^\sig{}_\nu \partial_q^\mu \partial_q^\nu-\tilde U\right) \right]^n\De \,.
\end{align}
This expansion has proved extremely useful in many EFT scenarios, including \cite{Henning:2014wua,Drozd:2015rsp,Filoche:2022dxl,Larue:2023uyv,Larue:2025yar}.
Yet, a truncation at any order is unable to capture non-perturbative effects; it does not predict an imaginary part, missing any possible vacuum instabilities generated by the background field, including the Schwinger effect.
This is of course related to the intrinsic character of the expansion, which in general would be just asymptotic (and not convergent).
In order to recover non-perturbative effects from this series, a Borel resummation could be employed. However, it would require knowing all the terms in the series, whereas it is usually truncated for computational reasons.
To avoid these issues, here we will rather modify the expansion right from the beginning; the result, as we will show, will be well suited to tackle the strong-field case of interest.

\subsection{Strong-field {CDE} expansion}
Let us start from the Gaillard--Cheyette expression as written in Eq.~\eqref{eq:lagrangian_GC}.
We previously considered the hierarchy where the mass of the scalar field dominated over all  scales associated with the background fields.
Whenever we refer to strong-field  physics, we understand that one scale---here the one associated with the field strength---is now comparable to or even larger than the mass of the scalar. Other scales in the problem, corresponding to inhomogeneities of the field strength or associated to other background fields, are still considered to be small.

Ergo, instead of expanding in powers of the free propagator, we will now expand about an operator $\Q$ that includes the terms containing the field strength without derivatives\footnote{Explicitly, the additional term that we consider in the propagator comes from $\tilde F_{\mu\nu}\partial_q^\mu = \frac{1}{2}F_{\mu\nu}\partial_q^\mu+ \mathcal{O}(\partial)$, where $\mathcal{O}(\partial)$ indicates terms with at least one derivative acting on the field strength.}
\begin{align}
    \Q&\equiv\frac{1}{4}F_{\mu\nu}F^\nu{}_\rho\partial_q^\mu\partial_q^\rho + F_{\mu\nu} \,q^\mu \partial_q^\nu -q^2+m^2 \,. 
\end{align}
Similar ideas have been also developed in other setups---for instance a chemical potential was included in Ref.~\cite{Larue:2025yar}. Here, however, the presence of $\partial_q$ in the new operator $\Q$ makes it necessary to carefully recast the expansion. 

At this point, it proves convenient to introduce further notation to simplify the writing in the following. 
We are going to treat as perturbations the elements
\begin{align}\label{eq:perturbation}
  \begin{split}
    L&\equiv L_{\rm inhom}-\tilde U\,,\\
    L_{\rm inhom}&\equiv \left\{\delta \tilde F_{\nu\mu}\partial^\mu_q,\frac{1}{2}F^\nu{}_\rho\partial^\rho_q-q^\nu \right\} + \delta \tilde F_{\nu\mu} \delta \tilde F_{\nu\rho}\partial_q^{\mu\rho}\,,\\
    \delta \tilde F_{\mu\nu}&\equiv \sum_{k=1}^{\infty} \frac{(k+1)\mathi^k}{(k+2)!} \partial_{\alpha_1\cdots \alpha_k} F_{\mu\nu} \partial_q^{\alpha_1\cdots \alpha_k} \,,
  \end{split}
\end{align}
 and  $\{\cdot,\cdot\}$ denotes the usual anticommutator. 
 In this way, we arrive at a compact expression for the effective Lagrangian,
\begin{align}\label{eq:expansion}
 \begin{split}
 \mathcal{L}_{\rm eff}&=\mathi \int \frac{\dxd[q][\dime]}{(2\pi)^\dime} \int^{M^2} \dx[m^2]
 \frac{1}{-(\tilde F_{\nu\mu} \partial_q^\mu+ q_\nu )^2 +m^2+\tilde U  }
 \\
 &= \mathi \int \frac{\dxd[q][\dime]}{(2\pi)^\dime} \int^{M^2}\dx[m^2]
 \frac{1}{\Q - L }\colred{(q)}
 \\
 &= \mathi \int \frac{\dxd[q][\dime]}{(2\pi)^\dime} \int^{M^2} \dx[m^2]
 \sum_{n=0}^\infty \left[\left(\GQ* L\right)^n \iQ \right](q)\, ,
 \end{split}
\end{align}
where we have defined the Green function  of $\Q$ in momentum space ($\GQ$) and \colred{its integral ($\iQ$)} by the equations
\begin{align}\label{eq:def-iQ}
 \begin{split}
  \Q(q\colred{,\partial_q})\GQ(q,p)&=\delta^{(\dime)} (q-p)\,,
  \\
  \iQ(q):&=\colred{\int\frac{\dxd[p][\dime]}{(2\pi)^\dime}\, \GQ(q,p)  \quad\Leftrightarrow\quad \Q(q\,,\partial_q)\iQ(q)=1 }\,, \\
 \end{split}
\end{align}
as well as introduced the operation $*$ to explicitly indicate the convolution action of $\GQ$:
\begin{align}
 \left[\GQ* L\,\iQ\right](q)
 =\int \frac{\dxd[p][\dime]}{(2\pi)^\dime} \GQ(q,p) L(p,\partial_p)\iQ(p)\, .
\end{align}
\colred{A few comments are in order. Following the notation in the literature, the element in the second line of \eqref{eq:expansion} is not the Green function of the differential operator $\Q-L$ but, rather,  the function which when acted upon by $\Q-L$ gives 1. In this sense it is denoted $\frac{1}{\Q-L}(q)$, but it should not be confused with the inverse of the corresponding differential operator, i.e. the Green function $\bra{q}\frac{1}{\Q-L}\ket{p}$. It is also for this reason that we have introduced both $\iQ$ and $\GQ$ for the operator $\Q$: the former to satisfy the Gaillard--Cheyette ``inverse'' condition, and the latter to perform the  perturbative computation. In contrast, when expanding around the free theory, the free propagator ($-q^2+m^2$) formally accomplishes both roles. As a simple check, note that applying $(\Q - L)$ to the series in Eq.~\eqref{eq:expansion} indeed gives just a factor one.  In addition, it is important to observe that the second line {of \eqref{eq:def-iQ}} is simply the convolution of $\GQ$ with the unit function. So, if desired, one could write Eq.~\eqref{eq:expansion} using just $\GQ$, by making explicit the trivial integral.}

Lastly, it is important to observe that the $n=0$ term alone corresponds to the Heisenberg--Euler Lagrangian~\cite{Weisskopf:1936hya}, cf. Sec.~\ref{sec:n0} below. In effect, it already contains all the constant-field information, such as the vacuum instabilities causing the Schwinger effect, without having to go through a further resummation. As can be seen from Eqs.~\eqref{eq:perturbation} and \eqref{eq:expansion}, the method proposes a natural grading in the number of derivatives acting on the field strength.
Indeed, this is an improved, strong-field version of the CDE, in which corrections coming from small inhomogeneities of a strong electromagnetic field can be step by step added.

\subsection{Building blocks and Green function}
Let us derive an explicit expression for the elements introduced to expand the Gaillard--Cheyette Lagrangian, i.e $\iQ$ and $\GQ$.
First, $\iQ$ can be found in Ref.~\cite{Brown:1975bc}, where it is computed by injecting a given Ansatz in our Eq.~\eqref{eq:def-iQ} and solving the resulting system of differential equations. In our conventions, we obtain the simple propertime expression
\begin{align}\label{eq:iQ}
 \begin{split}
\iQ(q) = \int_0^\infty \dd s \,e^{-m^2 s}e^{qAq+C}\,, \hspace{1cm}
  \begin{cases}
 A(s)\equiv-F^{-1}\tanh Fs \\
 \displaystyle C(s)\equiv-\frac{1}{2} \tr \log \cosh Fs
 \end{cases}\,.
 \end{split}
\end{align}
Note that here and in the following, whenever  indices are not written explicitly, it means that they are contracted with the neighboring indices  matrix-wise, for instance $(F^2)^{\alpha\beta}=F^\alpha{}_\gamma F^{\gamma\beta}$, while the natural contractions are intended, 
\begin{align}
    qAq \equiv q^\alpha A_{\alpha\beta}q^\beta.
\end{align}

Next comes the derivation of the Green function $\GQ$.
Of course, there are several ways to accomplish this task. We are going to resort to the worldline formalism and develop it in momentum space, which provides an efficient way to compute $\GQ(q,p)$. An alternative would be to proceed as in the computation of $\iQ$ offered in Refs.~\cite{Franchino-Vinas:2023wea, Brown:1975bc} , considering a few necessary subtle modifications to the initial Ansatz, since we are not computing a simple Fourier transform and boundary conditions will also differ from those in the literature. 

Coming back to the worldline approach, by exponentiating the differential operator $\Q$ we can recast the problem into a path-integral,
\begin{align}\label{eq:Zwl}
 \begin{split}
 \GQ(q,p)&=\bra{q}\colred{\Q^{-1}}\ket{p}\\
 &=\int_0^\infty \dd s\, \bra{q}e^{-s\Q}\ket{p}\\
 &=\int_0^\infty \dd s\,e^{-m^2 s}\int_{k(0)=\colred{p}}^{k(s)=\colred{q}} \mathcal{D}k(t)\, e^{\colred{-}\int_0^s \dd t \,L_{WL}[k]}\\
 &\equiv\int_0^\infty \dd s\,e^{-m^2 s}Z_{WL}(q,p)\,.
 \end{split}
\end{align}
In this expression, \colred{we have factored out the contribution of the mass from $\mathcal{Q}$ and a Legendre transform gives the Lagrangian associated to the resulting Hamiltonian,}
\begin{align}\label{eq:Lwl}
    L_{WL}[k(t)] = \dot k(t) F^{-2} \dot k(t) - 2k(t) F^{-1} \dot k(t)\,,
\end{align}
and we have introduced the worldline partition function $Z_{WL}$, which we will now proceed to evaluate. Taking care of the non-trivial boundary conditions when integrating by parts, we get
\begin{align}
 \begin{split}
    Z_{WL}(q,p) &= \int_{k(0)=\colred{p}}^{k(s)=\colred{q}} \mathcal{D}k(t)\, e^{\colred{-}\int_0^s \dd t \,k \mathcal{O} k \colred{-}\left[k F^{-2} \dot{k}\right]_0^s}\, ,\\[10pt]
    \mathcal{O}_{\al\be}&\equiv-F^{-2}_{\al\be}\partial^2_t-2F^{-1}_{\al\be}\partial_t \,.
    \end{split}
\end{align}
At this step, an astute change of variables involves splitting the paths into a classical and a quantum contribution, $k(t)=k_{cl}(t)+\varepsilon(t)$. In this expression, the classical trajectory $k_{cl}(t)$  satisfies $\mathcal{O}_{\al\be}k^\be_{cl}(t)=0$ and the boundary conditions of the path-integral, such that the quantum fluctuations $\varepsilon$ satisfy instead Dirichlet boundary conditions.
The solution for the classical trajectory is
\colred{
\begin{align}
    k_{cl}(t)=\left(1-e^{-2Fs}\right)^{-1}\left(-e^{-2Ft}(q-p)+q-e^{-2Fs}p\right)\,,
\end{align}}
as can be straightforwardly checked by replacing into the defining equation and evaluating at the boundary points, namely $t=0$ and $t=s$.
The path-integral then factorizes in the following way, where only the factor in front of the integral depends on $p$ and $q$,
\begin{align}
 \begin{split}
    Z_{WL}(q,p)&= e^{\colred{-}\left[k_{cl} F^{-2} \dot{k}_{cl}\right]_0^s}\,\int_{\varepsilon(0)=0}^{\varepsilon(s)=0} \mathcal{D}\varepsilon(t)\, e^{\colred{-}\int_0^s \dd t \,\varepsilon \mathcal{O}\varepsilon}\,\,.
 \end{split}
\end{align}
Massaging a bit this result, we finally obtain a proper-time expression for the Green function
\begin{align}\label{eq:GQ}
 \begin{split}
 \GQ(q,p) &= \int_0^\infty \dd s \, a(s)e^{-m^2 s}e^{b(F,s,q,p)}\,,
 \\
    a(s)&\equiv \colred{-\mathi}(4\pi)^{d/2}\text{det}^{-1/2}(-F\sinh Fs)\,,
    \\
 b(s,q,p)&\equiv qRq+pRp+pSq\,,
 \\
 R(s)&\equiv-F^{-1}\coth Fs\,,
 \\
 S(s)&\equiv 4F^{-1}(1-e^{\colred{-}2Fs})^{-1}\,.
 \end{split}
\end{align}
A clever way of computing the coefficient $a$, which is given by the determinant of the operator $\mathcal{O}$ on the domain of functions satisfying Dirichlet boundary conditions, is to deduce it from the expression of $\iQ$ through the \colred{second line in Eq.~\eqref{eq:def-iQ}}, which simultaneously provides the consistency check $R-\frac{1}{4}S^TR^{-1}S=A$, viz. Eq.~\eqref{eq:iQ}.

This is the right moment to pause and comment on the difference between our approach and the expansion proposed in \cite{Gusynin:1998bt}. 
The approach in Ref.~\cite{Gusynin:1998bt} requires the coincident limit of the worldline path-integral in position space, while contributions from derivatives of the field strength are introduced with the help of an external source $\eta$ and the corresponding generating functional, $\tilde Z_{WL}[\eta](x,x)$. Instead, in this manuscript we use the  Green function in momentum space, which contains all the information regarding powers of the field strength and was computed using the (momentum-space) worldline method, cf. Eq.~\eqref{eq:Zwl}.

The usual motivation to introduce momentum space is to switch from a position-space derivative operator to a function of momentum in momentum space, which is evidently much simpler to invert. This is not the case in our expansion, since $\Q$ contains derivatives of the momenta; nevertheless, it turns out to be more efficient to consider the diagonal of the Green function (in position space) and Fourier transform than using a point-splitting in position space.

In effect, let us write the quadratic operator of the Lagrangian in position-space, which can be obtained either from $\Q-L$ or independently by setting the Fock--Schwinger gauge in $(D^2+m^2)$.
The operator of quantum fluctuations can be formally written as
\begin{align}\label{eq:operator_position_space}
    \left(\partial^2 -\mathi F_{\mu\nu}(x-x')^\nu \partial^\mu -\frac{1}{4}F_{\mu\rho}F^\rho{}_\nu (x-x')^\mu(x-x')^\nu \right) + H(\partial_x,(x-x'))+U(x)\, ,
\end{align}
where terms with derivatives of $F$ have been collected in $H$.
In its turn, the effective action can be obtained by considering the Green function of this operator, taking the coincidence limit and integrating over the mass and the whole spacetime.
An expansion in powers of $H$ and $U$ for the Green function is, of course, in principle possible; however, the dependence of Eq.~\eqref{eq:operator_position_space} on both variables $x$ and $x'$ renders the expressions cumbersome, as could have been expected.


Before carrying out  explicit computations of the effective Lagrangian in Sec.~\ref{sect:inhom}, we are going to present in the following some generalities regarding its initial terms, corresponding to $n=0$, $1$ and $2$ in Eq.~\eqref{eq:expansion}.


\subsection{Constant field part: $n=0$}\label{sec:n0}
The first term arising in the expansion Eq.~\eqref{eq:expansion}, which corresponds to the Heisenberg--Euler effective action, is easily computed once we have determined $\iQ$, given that it involves no propagator $\GQ$:
\begin{align}
 \begin{split}\label{eq:n0}
    \mathcal{L}_{HE} \equiv \mathcal{L}_{\mathrm{eff}}^{n=0}
    &= \mathi \int \frac{\dxd[q][\dime]}{(2\pi)^\dime} \int^{M^2} \dx[m^2] \iQ(q)\\
    &= \colred{-}\frac{\colred{1}}{(4\pi)^{d/2}}\int_0^\infty \frac{\dd s}{s^{1+d/2}} e^{-M^2 s}\text{det}^{1/2}\left(\frac{F s}{\sinh F s}\right)\\
    &\hspace{-2.5pt}\overset{d=4}{=}\colred{-} \int_0^\infty \frac{\dd s}{s} e^{-M^2 s}\frac{K_- K_+}{\sin sK_- \sinh sK_+} \,.
 \end{split}
\end{align}
Two remarks are in order for these expressions. First, the diagonalisation of $F$ in $d=4$ appearing in the last line is detailed in App.~\ref{app:diagoF}; the scalar quantities $K_{\pm}$ are defined in terms of the quadratic Lorentz invariants built from the electromagnetic field. On the other hand, the integrand in the second line of Eq.~\eqref{eq:n0}, which is valid in any dimension $d$, diverges for small $s$, rendering the integral ill-defined. 

These divergences correspond to the ultraviolet sector of the theory. In fact, we should perform a renormalisation of the theory, by introducing counterterms for the first terms in an expansion in the propertime $s$ about zero. The terms in such an expansion are related to the heat kernel expansion,  which guarantees the  covariance of the renormalisation~\cite{Vassilevich:2003xt}.  In particular, the counterterms can be chosen in such a way that the effective Lagrangian vanishes as the field strength tends to zero.

To be concrete, define the coefficients $a_n$ from the following expansion:
\begin{align}
    e^{-M^2 s}\text{det}^{1/2}\left(\frac{F s}{\sinh F s}\right) &\equiv \sum_{n=0}^\infty a_n s^{n}\,.
\end{align}
It can be shown, for example, that $a_0$ plays the role of a cosmological constant, that $a_1$ vanishes in this theory, while $a_2$ is proportional to $F_{\mu\nu}F^{\mu\nu}$.
The above-mentioned renormalisation can be implemented by introducing in the theory an ultraviolet cutoff $\Lambda$ with mass dimensions, and performing the necessary subtractions,
\begin{align}\label{eq:renormalized_L}
    \mathcal{L}_{\rm HE,  ren} &\equiv \mathcal{L}_{\rm HE} - \frac{\colred{1}}{(4\pi)^{d/2}}\int_{\Lambda^{-2}}^\infty \frac{\dd s}{s^{1+d/2}} \sum_{n=0}^{\lfloor d/2\rfloor +1 } a_{n} s^n \,,
\end{align}
where $\lfloor \cdot \rfloor$ is the floor function.
This Lagrangian, as desired, vanishes in absence of an electromagnetic background; it is also finite in the large-$\Lambda$ limit, as long as we are working in $d\leq 4$ dimensions or with a homogeneous background field.

As a last comment, depending on the actual values of $K_{\pm}$, the integrand in Eq.~\eqref{eq:n0} will  develop poles whenever the matrix $\sinh (Fs)$ possesses imaginary eigenvalues. Restricting us to the four-dimensional case, depending on the actual values of $K_{\pm}$, the integrand might become singular for $s=n\pi/K_{\pm}$, $n=1,2,\cdots$. 
In order to obtain a well-posed integral, we can give a prescription to circumvent the poles in the complex $s$-plane, with the result that the effective Lagrangian acquires an imaginary part. This can be understood in terms of the Schwinger effect, which signals the instability of the vacuum and will be further discussed in Sec.~\ref{sec:Schwinger}. 


\subsection{Pre-computing the $n=1$ term} 
We will now go through the computation of the $n=1$  term of Eq.~\eqref{eq:expansion}, i.e. the contribution containing one propagator $\GQ$.  Even though the explicit form of the operator $L$ is known in our theory, cf. Eq.~\eqref{eq:perturbation}, we will keep our discussion in the following as general as possible by working with an arbitrary polynomial in the variable  $q$ and its associated derivatives $\partial_q$. The first-order contribution is given by
\begin{align}
 \begin{split}
 \mathcal{L}^{n=1}_{\rm eff}&= \mathi\int \frac{\dxd[q][\dime]}{(2\pi)^\dime} \int^{M^2} \toL\dx[m^2] \,\GQ* L\,\iQ\\
 &=\mathi\int_0^\infty \toL\dd s\,\dd u\int \frac{\dxd[q][\dime]}{(2\pi)^\dime} \frac{\dxd[p][\dime]}{(2\pi)^\dime} \int^{M^2} \toL\dx[m^2] e^{-m^2(s+u)}a(s)
 e^{b(F,s,q,p)}L(p,\partial_p)e^{p A(u)p +C(u)} \, ,
 \end{split}
 \end{align}
 where we shall recall that $A(s)=-F^{-1}\tanh Fs$, from Eq.~\eqref{eq:iQ}, and  we have utilized the expressions for the propagator which were obtained in momentum space. 
 In addition, the derivatives contained in $L$ are intended to act on the right and will give rise to a simple polynomial in $p$, which we will call $P$; the computation is straightforward and, after the integration over the auxiliary mass, one obtains
 \begin{align}
     \begin{split}
  \mathcal{L}^{n=1}_{\rm eff}&=\mathi\int_0^\infty \dd s\,\dd u \frac{-e^{-M^2(s+u)}}{s+u} a(s) \int \frac{\dxd[q][\dime]}{(2\pi)^\dime}\frac{\dxd[p][\dime]}{(2\pi)^\dime}\, P(p)e^{b(F,s,q,p)+p A(u)p +C(u)} \,.
   \end{split}
 \end{align}
In order to further massage  this expression, we introduce a source $B$ for $p$, such that we can replace the momenta in $P(p)$ by $\partial_B$ and evaluate simple Gaussian integrals: 
 \begin{align}
     \begin{split} \label{eq:n1}
&  \mathcal{L}^{n=1}_{\rm eff}
 =\mathi\int_0^\infty \frac{\dd\tau}{\tau} e^{-M^2\tau} \int_0^\tau\dd u\,
 a(s) \left. P(\partial_B) \int \frac{\dxd[q][\dime]}{(2\pi)^\dime}\frac{\dxd[p][\dime]}{(2\pi)^\dime} e^{b(F,s,q,p)+p A(u)p +C(u) +B\cdot p}\right\vert_{B=0}\\
 &=\frac{\colred{-1}}{(4\pi)^{d/2}}\int_0^\infty \frac{\dd\tau}{\tau^{1+d/2}} e^{-M^2\tau}\text{det}^{1/2}\left(\frac{F\tau}{\sinh F\tau}\right) \int_0^\tau\dd u\,
 \left.P(\partial_B) e^{-\frac{1}{4}B[A(\tau-u)+A(u)]^{-1}B}\right\vert_{B=0}\, .
 \end{split}
\end{align}
In order to simplify the notation, we have introduced a total propertime in this expression, which will prove useful also in the discussion of the higher-$n$ terms:
\begin{align}
    \tau\equiv s+u.
\end{align}
We note that the only effect of the perturbation $L$ on the result is through the polynomial $P(\partial_B)$ in the last line of Eq.~\eqref{eq:n1}. The main difference with the $n=0$ contribution is that here we have a further propertime integral, which somehow masks the presence of singularities. Note that we do not expect to be able to explicitly compute the integral over $\tau$, as it contains terms similar to those in the $n=0$ case. The $u$ integral, on the other hand, resembles the structure that has been integrated in Ref.~\cite{Gusynin:1998bt}.  Using the specific results obtained for the quantities involved in our scenario,  an explicit computation will be done in Sec. \ref{sect:inhom}, trying to be left with just a single integral, whose analytic structure can be readily identified. 

Nevertheless, at this point we can already observe that ultraviolet divergences are in general present in the $n=1$ terms for $d>4$ and reveal themselves also  in this case as divergences for small $\tau$. Indeed, one should proceed as previously done around Eq.~\eqref{eq:renormalized_L}, resulting in the renormalisation of terms which involve derivatives of the field strength. 
Even though in the following we are not going to further worry about the renormalisation of the theory, this divergent behaviour is more general, since ultraviolet divergences may appear for arbitrary $n$, depending on dimension of the underlying spacetime and the properties of the background field. In any case, it can be readily shown that the degree of divergence decreases as $n$ increases.


\subsection{Pre-computing  the $n=2$ term}
The computation at the $n=2$ order involves two propagators, which can be \emph{mutatis mutandis} introduced using the corresponding momentum-space representations:
\begin{align}
 \begin{split}
 \mathcal{L}^{n=2}_{\rm eff}&= \mathi\int \frac{\dxd[q][\dime]}{(2\pi)^\dime} \int^{M^2} \toL\dx[m^2] \,\GQ*L\,\GQ* L\,\iQ\\
 &=\mathi\int_0^\infty \toL\dd s\,\dd u\,\dd v\int \frac{\dxd[q][\dime]}{(2\pi)^\dime} \frac{\dxd[p][\dime]}{(2\pi)^\dime} \frac{\dxd[k][\dime]}{(2\pi)^\dime} \int^{M^2} \toL\dx[m^2] e^{-m^2(s+u+v)}\\
 &\hspace{4cm} \times a(s)e^{b(s,q,p)}L(p,\partial_p)a(u)e^{b(u,p,k)}L(k,\partial_k)e^{k A(v)k +C(v)}\\[5pt]
 &=\mathi\int_0^\infty \toL\dd s\,\dd u\,\dd v
\frac{-e^{-M^2(s+u+v)}}{s+u+v}a(s) a(u) \int \frac{\dxd[q][\dime]}{(2\pi)^\dime} \frac{\dxd[p][\dime]}{(2\pi)^\dime} \frac{\dxd[k][\dime]}{(2\pi)^\dime}  \\
 &\hspace{5cm}\times e^{b(s,q,p)}P_1(p,k)e^{b(u,p,k)}P_2(k)e^{k A(v)k +C(v)}\, .
 \end{split}
 \end{align}
 The chief novelty with respect to the $n=1$ case is that now we have two polynomials $P_i$, as a consequence of the two $L$-insertions present in the expansion.
 Introducing once more the total propertime variable
 \begin{align}
     \tau=s+u+v,
 \end{align}
 as well as sources $B_1$ and $B_2$ respectively for the momenta $k$ and $p$, we are led to a reduced formula,
 \begin{align}\label{eq:n2}
     \begin{split}    
 \mathcal{L}^{n=2}_{\rm eff}&=\mathi\int_0^\infty \dd \tau \frac{e^{-M^2\tau}}{\tau} \int_0^\tau \dd u\int_0^{\tau-u} \toL\dd v \,
  a(s) a(u) \Big[P_1(\partial_{B_1},\partial_{B_2})P_2(\partial_{B_2})\\
 &\hspace{0.8cm}\times\int \frac{\dxd[q][\dime]}{(2\pi)^\dime} \frac{\dxd[p][\dime]}{(2\pi)^\dime} \frac{\dxd[k][\dime]}{(2\pi)^\dime} e^{b(F,s,q,p)+b(F,u,p,k)+k A(v)k +C(v)+B_1\cdot p+B_2\cdot k}\Big]_{B_1,B_2=0}
 \\
 &=\frac{\colred{-1}}{(4\pi)^{d/2}}\int_0^\infty \dd \tau \frac{e^{-M^2\tau}}{\tau^{1+d/2}} \text{det}^{1/2} \left(\frac{F\tau}{\sinh F\tau}\right) \int_0^\tau \dd u\int_0^{\tau-u} \toL\dd v\\
 & \hspace{1.5cm}\times\Big[P_1(\partial_{B_1},\partial_{B_2})P_2(\partial_{B_2}) 
 e^{-\frac{1}{4}(B_2+B_1 Y)W_2(B_2+B_1 Y)-\frac{1}{4}B_1 W_1 B_1}\Big]_{B_1,B_2=0} \,,
 \end{split}
\end{align}
where we have defined the following tensorial quantities:
\begin{align}
\begin{split}\label{eq:w12_y}
    W_1 &\equiv-F^{-1}\left[\tanh(sF)+\coth(uF)\right]\, ,\\
    W_2&\equiv-F^{-1}\left[\tanh\big((s+u)F\big)+\coth(vF)\right]\, ,
    \\
    Y&\equiv\frac{\cosh(sF)}{\cosh\big((s+u)F\big)} e^{uF}\,.
\end{split}
\end{align}
Note that, once again, only the expressions of $P_{1,2}$ depend on $L$ and once it is specified we are only left with the evaluation of the $(u,v)$ propertime integrals. Indeed, at this point it should be clear that the computation at arbitrary $n$ can be recast in a master formula consisting of $n$ polynomials $P_i$, which are functions of the sources derivatives $\partial_{B_i}$, while there would remain $n$ propertime integrals to perform. Of course, the complexity of these integrals greatly enhances with $n$.

Finally, let us also comment on the appearance of the $\text{det}{\,\sinh F\tau}$ prefactor, which is also present in the Heisenberg--Euler Lagrangian. In this approach, it emerges from the combination of the determinants of the momentum integrals with the several $a$ and $C$ factors present in the first line of Eq.~\eqref{eq:n2}.   
This factorisation appeared naturally in the worldline expansion \cite{Gusynin:1998bt} and can be likewise generalized to any $n$ in the present framework. Moreover, it was recently conjectured that this is the only contribution containing invariants made of powers of the field strength (and not their derivatives)~\cite{Navarro-Salas:2020oew}.  In our current discussion,  this can be seen only as a computational remark instead of a physical one, as the prefactor could be modified by successive application of the $P_i$ operators on the exponential terms. A proof should instead put constraints on the factors accompanying this determinant, as has been performed in Refs.~\cite{Franchino-Vinas:2023wea, Franchino-Vinas:2025ejo}.


\section{The one-scale resummed effective field theory}\label{sect:inhom}

In this section we will come back to Eq.~\eqref{eq:perturbation}, focusing our computations on an electromagnetic background. Thus, we set $U=0$ and analyse the corrections to the Heisenberg--Euler Lagrangian  coming from (small) derivatives of $F$.
To mention a few applications of these contributions, recently it has been shown that the first derivative terms are required to study polarisation flip in circularly-polarized two-to-two photon scattering at the perturbative level \cite{Heinzl:2025xye}, and can be incorporated in the investigation of the Schwinger pair creation~\cite{Franchino-Vinas:2024jvc, Karbstein:2021obd}.

Contrary to the discussion in those references, we will be interested in a derivative expansion, which, at a given order $j$, contains all the contributions with $j$ derivatives acting on the background fields. In modern language, this can be seen as an EFT with two different \colred{types of scales arising from the background, which have to be compared with the mass scale: one is set by powers of the field strength, schematically $\left(\frac{F}{M^2}\right)^i$, while the other corresponds to derivatives $\left(\frac{\partial}{M}\right)^i$ acting on the field strength. The latter play the role of an ordinary EFT scale in our multiscale analysis, i.e. the effective action is an expansion in  powers of this scale, and its associated operators will be called ``basic operators''. In contrast, we are going to assume that the former scale could be of order one, meaning that they can not be simply considered in a perturbative fashion. By taking them into account to all-order, they are going to appear as resummed (tensorial) form factors dressing the basic operators. From the standard EFT perspective, they correspond to a resummation of a tower of operators; Instead from} a  Wilsonian perspective, they could be considered as generalized runnings, setting a secondary scale for each single basic operator present in the theory. 

{\colred
In formulae, we are dealing with an expansion of the form
\begin{align}
\begin{split}
    \mathcal{L}_{\rm eff}&= M^{\dime} \sum_{N=0}^{\infty} \sum_j g_{N,j}\left(\frac{F}{M^2}\right)\mathcal{O}_{N,j}\,,\\
    \mathcal{O}_{N,j}&\sim \frac{1}{M^{2k+N}}\partial^{i_1}F\dots\partial^{i_k}F\,,\quad \text{with }\sum_k i_k = N\,,
\end{split}
\end{align}
where $g_{N,j}$ are tensorial functions (form factors) and the basic operator $\mathcal{O}_{N,j}$ displays two indices: $N$ labels the number of derivatives, while  the multi-index $j$ hides the tensorial details and serves to label the different operators. To be more explicit about the computation at order $N$, consider \eqref{eq:expansion} together with the general perturbation \eqref{eq:perturbation}. On the one hand, one can drop the higher orders in $\delta\tilde F$, i.e. truncate the sum at $k_{\rm max} = N$ in Eq.~\eqref{eq:perturbation}.
On the other hand, since the perturbation $L$ contains a sum of terms, each involving at least one derivative, the expansion in Eq.~\eqref{eq:expansion} can be truncated at $n_{\rm max} = N$. The expression resulting from this double truncation contains all the necessary terms for the computation at level $N$.
This shows that we can reach any desired order $N$, the only price being that the number of terms grows as $N^2$.
}


\subsection{Basic operators at second order in derivatives}
The lowest corrections to the Heisenberg--Euler effective action \colred{($N=0$)} come from terms with \colred{$N=2$} derivatives acting on powers of the field strength $F$. Such contributions were considered in Ref.~\cite{Lee:1989vh}, though they corresponded just to a specific background. A similar idea can be found instead in Ref.~\cite{Gusynin:1998bt}, where the worldline method was employed for the computation. 
In our calculation, \colred{as discussed in the previous section, we can truncate the expansion in Eq.~\eqref{eq:expansion} for the effective Lagrangian by limiting the number of derivatives; indeed, according to Eq.~\eqref{eq:perturbation}, this amounts to considering the contributions from just  a finite number of terms, since the expression for $\delta \tilde{F}_{\mu\nu}$ is an expansion in the number of derivatives acting on a single field strength.} The relevant perturbations in this section are
\begin{align}
\begin{split}\label{eq:inhom_perturb}
    \de\tilde F_{\mu\nu}&=\frac{i}{3}\partial_{\al}F_{\nu\mu}\partial_q^{\al} -\frac{1}{8}\partial_\sig\partial_\lambda F_{\nu\mu}\partial_q^{\sig\lambda}+ \mathcal{O}(\partial^3) \,,
    \\
     L_{\textrm{inhom}}&= \frac{i}{3}\left(F^\nu{}_\rho \partial_q^{\al\mu\rho}-2q^\nu\partial_q^{\al\mu}-\eta^{\al\nu}\partial_q^\mu\right)\partial_\al F_{\nu\mu}
     -\frac{1}{9}\partial_\al F_{\nu\mu}\partial_\be F^\nu{}_\rho\partial_q^{\al\mu\be\rho}
     \\
     &\hspace{0.3cm}-\frac{1}{8}\left(F^\nu{}_\rho\partial_q^{\sig\lambda\mu\rho}-q^\nu\partial_q^{\sig\lambda\mu}-\eta^{\sig\nu}\partial_q^{\lambda\mu}-\eta^{\lambda\nu}\partial_q^{\sig\mu}-\eta^{\mu\nu}\partial_q^{\sig\lambda}\right)\partial_\sig\partial_\lambda F_{\nu\mu}+ \mathcal{O}(\partial^3)
     \,.
\end{split}
\end{align}

In the following, we will show how to get the expressions of the form factors $y_1$ and $y_2$ appearing in
\begin{align}
 S_{\mathrm{eff}} &= \int {\rm d}^4x\,\mathcal{L}_{HE}(F) + \int {\rm d}^4x\, \mathcal{L}_{\rm eff}^{\partial^2} +\mathcal{O}(\partial^4) \,,
 \\
  \mathcal{L}_{\rm eff}^{\partial^2} &\equiv y_1^{\al\be\mu\nu}(F) \,\partial_\al\partial_\be F_{\mu\nu}+y_2^{\al\be\mu\nu\rho\sig}(F) \,\partial_\al F_{\mu\nu} \partial_\be F_{\rho\sig} \,.
  \label{eq:goal}
\end{align}
At $n=1$ in our expansion Eq.~\eqref{eq:expansion}, we consider elements of $L_{\rm inhom}$ which contain two derivatives of $F$, either as 
$\partial^2F$ or $\partial F\partial F$,
while for $n=2$, we keep elements of $L_{\rm inhom}$ which contain one derivative of $F$, since they will appear ``squared''.
Using the pre-computed expressions~\eqref{eq:n1} and \eqref{eq:n2}, we immediately get an expression of the form\footnote{Note that we could have integrated by parts in order to include $\mathcal{P}_1$ directly in $\mathcal{P}_2$. However, we prefer it to keep it for two reasons. First, integrating by parts one would act also on the prefactor, thus unnecessary complicating the computation. Second, it allows a direct comparison with the available literature.}
\begin{align}\label{eq:inject}
  \begin{split}
    \mathcal{L}_{\rm eff}^{\partial^2} &=\frac{\colred{-1}}{(4\pi)^{d/2}}
    \int_0^\infty \frac{\dd\tau}{\tau^{1+d/2}} e^{-M^2\tau}\text{det}^{1/2}\left(\frac{F\tau}{\sinh F\tau}\right)\\
    \times&\int_0^\tau\dd u \bigg[ \mathcal{P}_1^{\al\be\mu\nu}\partial_\al\partial_\be F_{\mu\nu}
    +\mathcal{P}_2^{\al\be\mu\nu\rho\sig}\partial_\al F_{\mu\nu}\partial_\be F_{\rho\sig}+ \int_0^{\tau-u} \toL \dd v\,\mathcal{P}_3^{\al\be\mu\nu\rho\sig}\partial_\al F_{\mu\nu}\partial_\be F_{\rho\sig}\bigg] \,,
  \end{split}
\end{align}
\colred{which defines the pre-form factors $\mathcal{P}_i$.}
Inasmuch as the expressions easily become tedious---around $10^3$ terms with different tensorial structures are present for $n=2$---we have implemented a code using the xAct package in Wolfram Mathematica~\cite{xact}. We will now describe its mechanism, starting from the expressions of $\mathcal{P}_i$ computed from $L_{\textrm{inhom}}$, which are functions of $F$ and the propertime parameters, to afterwards perform the $(u,v)$ integrals and finally simplify as much as possible the resulting form factors.


\subsection{Procedure: a simple example}\label{sect:example}
It is useful to work out by hand the calculations corresponding to  a simple piece of the perturbation $L$, in order to understand the mechanics which is going to guide the full computation. Let us examine the $n=2$ contribution of the term with the fewest momentum derivatives in Eq.~\eqref{eq:inhom_perturb}, i.e.
\begin{align}
    L_{\rm ex}(q,\partial_q)\equiv  \partial^\mu F_{\al\mu}\partial_q^\al \,.
\end{align}
To compute the effective Lagrangian corresponding to this contribution, namely the analogue of Eq.~\eqref{eq:n2} in this simple case, we must first obtain the expressions of the polynomials $P_1$ and $P_2$, by applying $L$ respectively on $\GQ$ and $\iQ$. We obtain
\begin{align}
 \begin{split}
    P_1(p,k) e^{b(F,u,p,k)}&=L(p,\partial_p)e^{pR(u)p +kR(u)k +kS(u)p}\\
    &\hspace{2cm}\implies P_1(p,k)= \left[ 2p_\rho R^{\rho\al}(u)+k_\rho S^{\rho\al}(u)\right] \,\partial^\mu F_{\al\mu}\,, 
    \\[10pt]
    P_2(k) e^{kA(v)k+C(v)}&=L(k,\partial_k) e^{kA(v)k+C(v)}\implies P_2(k)= 2k_\sig A^{\sig\be}(v) \,\partial^\nu F_{\be\nu}\,. 
 \end{split}
\end{align}
Using their explicit form, we can now evaluate the equivalent of Eq.~\eqref{eq:n2} and get
\begin{align}\label{eq:Leff_ex}
    \mathcal{L}^{(n=2)}_{\rm ex}=\frac{\colred{-1}}{(4\pi)^{d/2}}\int_0^\infty \dd \tau \frac{e^{-M^2\tau}}{\tau^{1+d/2}} \text{det}^{1/2} \left(\frac{F\tau}{\sinh F\tau}\right) \int_0^\tau \dd u\int_0^{\tau-u} \toL\dd v \,\mathcal{P}_{\al\be}\,\partial_\mu F^{\al\mu} \partial_\nu F^{\be\nu}\,,
\end{align}
where we have defined the pre-form factor $\mathcal{P}_{\al\be}$ as expected,
\begin{align}
 \begin{split}
    \mathcal{P}_{\al\be} \,\partial_\mu F^{\al\mu} \partial_\nu F^{\be\nu}
    &\equiv \Big[P_1(\partial_{B_1},\partial_{B_2})P_2(\partial_{B_2}) e^{-\frac{1}{4}(B_2+B_1 Y)W_2(B_2+B_1 Y)-\frac{1}{4}B_1 W_1 B_1}\Big]_{B_1,B_2=0}\,,
    \end{split}
    \end{align}
    and a straightforward computation, using the definitions in Eq.~\eqref{eq:w12_y}, gives
    \begin{align}\label{eq:simple}
    \begin{split}   
    \mathcal{P}_{\al\be} &=-2\left(A(v)W_2Y^TR(u)\right)_{\be\al} -\left(A(v)W_2S(u)\right)_{\be\al}\\
    &=-2\left(\frac{e^{-uF}\sinh(sF)\sinh(vF)}{F \sinh(\tau F)} \right)_{\al\be}\,.
 \end{split}
\end{align}

Note that all the integrals present in Eq.~\eqref{eq:Leff_ex} are well-defined, though only when considering the full expression and not the individual terms of the sum present in Eq.~\eqref{eq:simple}. This is evidenced by taking the weak-field expansions of such terms: the lowest order contributions individually give a divergent piece, proportional to $2v/(u\tau F^2)$, but cancel each other in the sum.
This observation will be essential when dealing with the full expression, as it means that in general we cannot integrate individual terms without previous simplifications or groupings. Additionally, only the symmetric part of $\mathcal{P}_{\alpha\beta}$ will survive the contraction with $\partial_\mu F^{\al\mu} \partial_\nu F^{\be\nu}$. Even though the antisymmetric part does not prevent the integration in this simple example, a similar symmetrisation will prove crucial in the full case. Denoting the idempotent symmetrisation (anti-symmetrisation) by parenthesis (brackets) enclosing the corresponding indices, we have
\begin{align}
    \mathcal{P}_{(\al\be)} &=-2\left(\frac{\cosh(uF)\sinh(sF)\sinh(vF)}{F \sinh(\tau F)} \right)_{\al\be} \,,
\end{align}
which can be easily integrated in the propertimes $u$ and $v$:
\begin{align}
    \int_0^\tau \dd u\int_0^{\tau-u} \toL\dd v \,\mathcal{P}_{(\alpha\beta)}
    =\tau\left(\frac{1-\tau F \coth(\tau F)}{2 F^2}\right)_{\alpha\beta}\,.
\end{align}


\subsection{Weak-field expansion of the Lagrangian}
Given the large number of terms we are going to deal with when computing the effective Lagrangian, we found it instructive to validate our results by performing comparisons with those available in the literature at a few stages of the calculation. An extremely useful validity check of our computation is to consider the weak-field limit of $\mathcal{L}_{\rm eff}^{\partial^2}$, which corresponds also to its large-mass expansion. 
One benefit of such expansion, is that, at the level of Eq.~\eqref{eq:inject}, the tensorial structure greatly simplifies and the integrals in the propertime become trivial to evaluate.

When considering this weak-field expansion, we should express the result in a (minimal) basis, taking into account the above-mentioned  symmetries in the indices, the Bianchi identity and, possibly, integration by parts. Note that the only terms relevant for the present discussion are those with two derivatives, acting either on the same or on different field strengths. At mass dimension 6, the result can always be reduced to $\partial_\al F_{\mu\nu}\partial^\al F^{\mu\nu}$, while at mass dimension 10, a basis is made of 7 elements. It is important to mention that this number does not coincide with the one considered in usual Hilbert series, where onshellness is used to further reduce the basis~\cite{Henning:2017fpj}. In any case, an algorithm to reduce to a basis is described in Ref.~\cite{Muller:1996cq} and one suitable choice is given in App.~\ref{sec:minimal_basis}.

Let us note that, in performing the weak-field expansion, a certain grouping must be made in order to be able to integrate the pre-form factors $\mathcal{P}_i$.
Indeed, as we saw in Sec.~\ref{sect:example}, some of the involved terms possess divergences characterised by negative powers of $F$, i.e., of the form $\Phi(F)^{\al\be} = c_{-n}(\tau,u,v) (F^{-n})^{\al\be}+\cdots$ where $c_{-n}$ is not integrable for small propertimes. However, once the full symmetries are enforced on the indices of the pre-form factors, the divergent contributions cancel and the total result is well-defined.
Rather than being an inconvenience, these cancellations appearing in the weak-field approximation become a powerful tool to spot  terms that, grouped together in the resummed expression, are expected to simplify. This is very much like the cancellation between the two terms in Eq.~\eqref{eq:simple}; at the resummed level, this naive observation significantly increases the speed of the simplifications.
Finally, we compute all integrals in the weak-field expansion and obtain the coefficients of the operators up to mass dimension 10 in the minimal basis. 

As a validation of our approach, we have also computed the effective Lagrangian for scalar QED at this order employing the usual perturbative CDE, which is described in Sec.~\ref{CDE}. In the literature,  the result can be found for instance in Ref.~\cite{Fliegner:1997rk}, after all the intrinsic non-abelian contributions are dismissed. Once written in the same basis, our results are in agreement with Ref.~\cite{Fliegner:1997rk}; for the sake of completeness, our expression is provided in Eqs.~\eqref{eq:O3} and \eqref{eq:O5}. 
However, as pointed out already in Ref.~\cite{Franchino-Vinas:2023wea}, the strong-field derivative expansion obtained in Ref.~\cite{Gusynin:1998bt} does not lead to the correct weak-field expansion, even if the general approach in that article seems correct.


\subsection{Structure of the result and simplifications}
Before trying to brute-force integrate the expressions in the strong-field regime, let us focus on the tensorial structure inside the pre-form factors $\mathcal{P}_i$. In particular, we will be interested in finding a minimal basis of basic operators, as done in the weak-field case.

Taking advantage of the appropriate symmetrisations and the Bianchi identity, we can indeed reduce $\mathcal{P}_1$ to a sum of terms which have just $\Phi_1^{\mu\al}\Phi_2^{\nu\be}$ as their tensorial structure. In this expression and in the following, $\Phi_j$ simply represents a function of $F$ and its subindex $j$ is not a tensorial one: it is just a labelling. Note also that we keep the indices of $\partial_\al\partial_\be F_{\mu\nu}$ in Eq.~\eqref{eq:inject} fixed when adding several contributions in the pre-form factor. Similarly, $\mathcal{P}^{\al\be\mu\nu\rho\sig}_{2}$ and $\mathcal{P}^{\al\be\mu\nu\rho\sig}_{3}$ respectively consist of a sum of terms that can be recast into the form $\Phi_1^{\al\mu}\Phi_2^{\be\rho}\Phi_3^{\nu\sig}$ and\footnote{Even if slightly pedantic, we prefer to indicate both alternatives, which correspond just to a permutation in a pair of indices, in order to make it clear that they correspond to different tensorial structures once contracted as in Eq.~\eqref{eq:inject}. A similar comment is also valid  for the second and third line in Eq.~\eqref{eq:doubling}.} $\Phi_1^{\al\rho}\Phi_2^{\be\mu}\Phi_3^{\nu\sig}$.

To observe the symmetrisations that should be taken into account, consider the basic operator $\partial_\al\partial_\be F_{\mu\nu}$. The corresponding form factor will compulsory be symmetric in $(\al,\be)$, while antisymmetric in $(\mu,\nu)$; therefore, an expression as $\Phi_1^{\mu\al}\Phi_2^{\nu\be}+\Phi_2^{\mu\al}\Phi_1^{\nu\be}$ will cancel once contracted with the basic operator.
Thus we must enforce a few symmetries for the pre-form factors prior to their integration to get rid of divergent terms, to wit
\begin{align}\label{eq:doubling}
    \begin{split}
        \Phi_1^{\mu\al}\Phi_2^{\nu\be}&\rightarrow  \Phi_{[1}^{\mu\al}\Phi_{2]}^{\nu\be}\,,\\
        \Phi_1^{\al\mu}(F)\Phi_2^{\be\rho}(F)\Phi_3^{\nu\sig}(F)&\rightarrow  \frac{1}{2}\left(\Phi_1^{\al\mu}(F)\Phi_2^{\be\rho}(F)\Phi_3^{\nu\sig}(F) + \Phi_2^{\al\mu}(F)\Phi_1^{\be\rho}(F)\Phi_3^{\nu\sig}(-F)\right)\,,\\
        \Phi_1^{\al\rho}(F)\Phi_2^{\be\mu}(F)\Phi_3^{\nu\sig}(F)&\rightarrow \frac{1}{2}
        \left( \Phi_1^{\al\rho}(F)\Phi_2^{\be\mu}(F)\Phi_3^{\nu\sig}(F) + \Phi_2^{\al\rho}(F)\Phi_1^{\be\mu}(F)\Phi_3^{\nu\sig}(-F)\right)\,.
    \end{split}
\end{align}
Let us define the even and odd parts of a function as
\begin{align}
    \Phi_{\textrm{even}}(F)=\frac{\Phi(F)+\Phi(-F)}{2}\,,\quad
    \Phi_{\textrm{odd}}(F)=\frac{\Phi(F)-\Phi(-F)}{2}\,.
\end{align}
As a consequence of Furry's theorem, only the odd part of $y_1$ and the even part of $y_2$ in Eq.~\eqref{eq:goal} should contribute. We will thus keep only these terms and perturbatively check that the remaining contributions vanish (we have done this up to order 12 in the mass).

On top of these properties, we can introduce the diagonalisation of $F$ which is available in four dimensions, see Refs.~\cite{Batalin:1971au, Lee:1989vh, Gusynin:1998bt} and App.~\ref{app:diagoF} for the complete formulae. This can be readily used to separate the functional dependence from the tensorial structure, the latter being independent of the integration variables.
Essentially, this means that we can recast a function of the field strength as
\begin{align}\label{eq:F_diagonalisation}
    \Phi(F)^{\al\be}=\sum_i \Phi(f_i) A_{(i)}^{\al\be}\,,
\end{align}
where the $A_{(i)}$ can be easily understood when expressed in the language of diagonalisation of a matrix; introducing the change of basis matrix $P$,
\begin{align*}
    F=P\cdot \text{diag}(f_1,...,f_d)\cdot P^{-1}=\sum_k f_k\,P\cdot\text{diag}(\de_{jk})\cdot P^{-1}=\sum_k f_k\,A_{(k)} \,.
\end{align*}
The eigenvalues $f_i$ are made of  the two invariants available in four dimensions~\cite{Gusynin:1998bt}, namely 
\begin{align}
\label{eq:mathF}    \F&\equiv-\frac{1}{4}F^{\mu\nu}F_{\mu\nu} \,,
\\
\label{eq:mathG} \G&\equiv\frac{1}{8}\eps^{\mu\nu\rho\sig}F_{\mu\nu}F_{\rho\sig}\, .
\end{align}
Altogether, for $\mathcal{P}_1$ we will need to integrate expressions of the form
\begin{align}
    \mathcal{P}_1^{\al\be\mu\nu}\rightarrow \sum_{\{\Phi_1,\Phi_2\}}\sum_{i,j}
    \frac{1}{2}\left[\Phi_1(f_i)\Phi_2(f_j)-\Phi_2(f_i)\Phi_1(f_j)\right]_{\textrm{odd}}A_{(i)}^{\mu\al}A_{(i)}^{\nu\be} \,,
\end{align}
while similar expressions can be written down for the remaining pre-form factors. Importantly, in these expressions the dependence on $(u,v)$ is only through the scalar functions $\Phi_i$.


\subsection{Full form factors for the basic operators at order $\partial^2$}\label{sect:result}
One of the  main results of this article is the expression of the quadratic derivative correction to the Heisenberg--Euler effective action. Making use of the previous discussions, we arrive at the expression  
\begin{align}\label{eq:result}
  \begin{split}
    \mathcal{L}_{\rm eff}^{\partial^2} =
    \frac{\colred{-1}}{(4\pi)^{d/2}}\int_0^\infty \frac{\dd\tau}{\tau^{1+d/2}} e^{-M^2\tau}\text{det}^{1/2}\left(\frac{F\tau}{\sinh F\tau}\right)
    \bigg[ \sum_{i,j}
    R^{\rm sym}_0(f_i,f_j)
    A_{(i)}^{\mu\al}A_{(j)}^{\nu\be}\partial_\al\partial_\be F_{\mu\nu}&\\
    +\sum_{i,j,k}\left(
    R^{\rm sym}_1(f_i,f_j,f_k)A_{(i)}^{\al\mu}A_{(j)}^{\be\rho}A_{(k)}^{\nu\sig}
    + R^{\rm sym}_2(f_i,f_j,f_k)A_{(i)}^{\al\rho}A_{(j)}^{\be\mu}A_{(k)}^{\nu\sig}\right)\partial_\al F_{\mu\nu}\partial_\be F_{\rho\sig}\bigg] \,,&
  \end{split}
\end{align}
containing just a single integral over propertimes.
The symmetrisation is the one discussed in Eq.~\eqref{eq:doubling}; after diagonalisation it becomes
\begin{align}\label{eq:sym}
\begin{split}
    R^{\text{sym}}_l(f_i,f_j)&=\frac{1}{2}\Big(R_l(f_i,f_j)-R_l(f_j,f_i)\Big)\, ,\\
    R^{\text{sym}}_l(f_i,f_j,f_k)&=\frac{1}{2}\Big(R_l(f_i,f_j,f_k)+R_l(f_j,f_i,-f_k)\Big)\, .
\end{split}
\end{align}
To further simplify the notation, we introduce the functions
\begin{align}
    H(x)=\frac{x \coth x - 1}{x^2} \,,\quad \ct_i=\coth(\tau f_i)\,,
\end{align}
so that we can provide the following explicit expressions of the functions $R_i$:
\begin{align}\label{eq:R0}
\begin{split}
    R_0(f_i,f_j)=\frac{1-\tau^2 f_i^2H(\tau f_j)}{2(f_i+f_j)} \tau H(\tau f_i) \,,
\end{split}
\end{align}
\begin{align}\label{eq:R1}
\begin{split}
    R_1(f_i,f_j,f_k)&=
    \frac{\tau}{4 \left(f_i+f_j\right)}
    \left[ \Big(\tau ^2 f_j \left(f_i+f_j\right)H\left(\tau  f_j\right)+1\Big)\frac{H\left(\tau 
   f_i\right)-H\left(\tau  f_k\right)}{f_i-f_k} \right.\\
   &\hspace{2.4cm}+\left. \Big(\tau ^2 f_i \left(f_i+f_j\right)H\left(\tau  f_i\right)+1\Big) \frac{H\left(\tau  f_j\right)-H\left(\tau 
   f_k\right)}{f_j+f_k}\;\right]\\
   &\hspace{0.5cm}-\frac{\tau^5}{4}f_i f_jH\left(\tau  f_i\right) H\left(\tau  f_j\right) H\left(\tau f_k\right) \,,
   \end{split}
\end{align}
\begin{align}
\begin{split}\label{eq:R2}
    &R_2(f_i,f_j,f_k)=\frac{1}{9}\Bigg[
    -\tau ^3 H\left(\tau  f_i\right) H\left(\tau  f_k\right) -\frac{4 \tau  \left(\tau ^2 f_i^2 H\left(\tau  f_i\right)+1\right) H\left(\tau f_j\right)}{f_i^2-f_j^2}\\
    &+\frac{\tau  \left(H\left(\tau  f_i\right)-H\left(\tau  f_k\right)\right)}{f_i^2-f_k^2}
    + \tau\left(\frac{2 \ct_i \ct_j }{f_i f_j} +\frac{\ct_j \ct_k}{2 f_j f_k}+\frac{\ct_i \ct_k}{2 f_k \left(f_i-f_j\right)}+\frac{9 \left(\ct_i \left(\ct_j-\ct_k\right)+\ct_j \ct_k-1\right)}{4 \left(f_i-f_j\right) \left(f_i-f_j+f_k\right)}\right)\\
    &+\frac{9 \ct_j}{8 f_i f_j f_k}-\frac{2 \ct_i}{f_i f_j^2} - \frac{2 \ct_j}{f_i^2 f_j} -\frac{\ct_i \left(\ct_j \ct_k+1\right)}{2 f_i f_k^2}-\frac{\ct_j \left(\ct_i \ct_k+1\right)}{2 f_j f_k^2} +\frac{\ct_j-\ct_i}{f_i f_j \left(f_i-f_j\right)}+\frac{\ct_i+\ct_j}{f_i f_j \left(f_i+f_j\right)}\\
    &+\frac{\ct_k-\ct_i}{4 f_i^2 \left(f_i-f_k\right)}-\frac{\ct_j+\ct_k}{4 f_j^2 \left(f_j+f_k\right)}+\frac{-\ct_i \left(\ct_j \ct_k+1\right)-\ct_j-\ct_k}{4 f_i f_j \left(f_i+f_j+f_k\right)}+\frac{\ct_i \left(1-\ct_j \ct_k\right)+\ct_j-\ct_k}{4 f_i f_j \left(f_i+f_j-f_k\right)}\\
    &+\frac{3 \ct_i \ct_j \ct_k+9 \ct_i-3 \ct_k-4\tau  \ct_i f_j \left(\ct_j+\ct_k\right)}{8 f_i f_k \left(f_i-f_j\right)}+\frac{3 \ct_i \left(\ct_j \ct_k-3\right)-3 \ct_k+4\tau  \ct_i \ct_j f_i}{8 f_j f_k \left(f_i-f_j\right)}\\
    &+\frac{-\ct_i \left(\ct_j \ct_k+1\right)+\ct_j+\ct_k}{4 f_j \left(f_i-f_j\right) \left(f_i-f_j-f_k\right)}+\frac{-\ct_i \left(\ct_j \ct_k+1\right)+\ct_j+\ct_k}{4 f_i \left(f_i-f_j\right) \left(f_i-f_j-f_k\right)}+\frac{\ct_j \left(\ct_i \ct_k-1\right)}{2 f_j \left(f_i-f_j\right) \left(f_i-f_k\right)}\\
    &+\frac{3 \left(\ct_i \left(\ct_j \ct_k-1\right)+\ct_j-\ct_k\right)}{8 f_i \left(f_i-f_j+f_k\right)^2}
    +\frac{3 \left(\ct_i \left(\ct_j \ct_k-1\right)+\ct_j-\ct_k\right)}{8 f_j \left(f_i-f_j+f_k\right)^2}-\frac{\ct_i \left(\ct_j \ct_k+1\right)}{2 f_i \left(f_i-f_j\right) \left(f_j+f_k\right)}\\
    &+\frac{3 \left(\ct_i \left(\ct_j \ct_k-1\right)+3 \left(\ct_j-\ct_k\right)\right)}{8 f_i \left(f_j-f_k\right)^2}
    +\frac{3 \left(\ct_i \left(\ct_j \ct_k+3\right)+\ct_j+3 \ct_k\right)}{8 f_j \left(f_i+f_k\right)^2}
    \\
    &+\frac{\ct_i \left(6 \tau  \ct_k f_j+\ct_j \left(\ct_k-6 \tau  f_j\right)-1\right)+\ct_j \left(1-6 \tau  \ct_k f_j\right)-\ct_k+6 \tau f_j}{8 f_i \left(f_i-f_j\right) \left(f_i-f_j+f_k\right)}\\
    &+\frac{\ct_j \left(1-12 \tau  \ct_k f_i\right)+\ct_i \left(\ct_j \left(\ct_k-12 \tau  f_i\right)+12 \tau  \ct_k f_i-1\right)-\ct_k+12 \tau  f_i}{8 f_j \left(f_i-f_j\right) \left(f_i-f_j+f_k\right)}\\
    &+\frac{\ct_i \left(6 \tau  \ct_k f_j-\ct_j \left(\ct_k-6 \tau  f_j\right)-1\right)+\ct_j \left(6 \tau  \ct_k f_j-1\right)-\ct_k+6 \tau  f_j}{4 (f_i^2-f_j^2) \left(f_i+f_k\right)}\\
    &+\frac{\ct_j \left(3 \tau  \ct_k f_i+1\right)+\ct_i \left(\ct_j \left(\ct_k-3 \tau  f_i\right)+3 \tau  \ct_k f_i-1\right)-\ct_k-3 \tau  f_i}{4 (f_i^2-f_j^2) \left(f_j-f_k\right)}
    \\
    &-\frac{f_j \left(\ct_i \left(\ct_j \ct_k+9\right)+\ct_j+9 \ct_k\right)}{8 f_i (f_i^2-f_j^2) \left(f_i+f_k\right)}+\frac{f_i \left(-\ct_j \left(\ct_i \ct_k+9\right) + \ct_i+9 \ct_k\right)}{8 f_j (f_j^2-f_i^2) \left(f_j-f_k\right)}+\frac{\ct_j f_j \left(\ct_i \ct_k-1\right)}{4 f_i^2 \left(f_i-f_j\right) \left(f_i-f_k\right)}\\
    &+\frac{3 f_j \left(\ct_i \left(1-\ct_j
   \ct_k\right)-\ct_j+\ct_k\right)}{4 f_i^2 \left(f_i-f_j\right)
   \left(f_i-f_j+f_k\right)}+\frac{3 f_i \left(\ct_i \left(1-\ct_j \ct_k\right)-\ct_j+\ct_k\right)}{4 f_j^2 \left(f_i-f_j\right) \left(f_i-f_j+f_k\right)}-\frac{\ct_i f_i \left(\ct_j \ct_k+1\right)}{4 f_j^2 \left(f_i-f_j\right) \left(f_j+f_k\right)}\\
    &+\frac{f_j \left(\ct_i \left(-6 \tau  \ct_k f_j+\ct_j \left(6 \tau  f_j-7 \ct_k\right)+7\right)+\ct_j \left(9-6 \tau  \ct_k f_j\right)-9 \ct_k+6 \tau  f_j\right)}{8 f_i (f_i^2-f_j^2) \left(f_j-f_k\right)}\\
    &+\frac{3 f_i \left(\ct_k \left(4 \tau  \ct_j f_i+3\right)+\ct_i \left(4 \tau  \ct_j f_i+4 \tau  \ct_k f_i+3\right)+4 \tau  f_i\right)-7 \ct_j f_i \left(\ct_i \ct_k+1\right)}{8 f_j (f_j^2-f_i^2) \left(f_i+f_k\right)}\\
    &+\frac{3 f_j^2 \left(\ct_i \left(1-\ct_j \ct_k\right)-3 \ct_j+3 \ct_k\right)}{4 f_i^2 (f_i^2-f_j^2) \left(f_j-f_k\right)}+\frac{f_j^2 \left(\ct_i \left(\ct_j \ct_k+1\right)+\ct_j+\ct_k\right)}{4 f_i^2 (f_i^2-f_j^2) \left(f_i+f_k\right)}\\
    &-\frac{3 f_i^2 \left(\ct_i \left(\ct_j \ct_k+3\right)+\ct_j+3 \ct_k\right)}{4 f_j^2 (f_j^2-f_i^2) \left(f_i+f_k\right)}+\frac{f_i^2 \left(\ct_i \left(\ct_j \ct_k-1\right)+\ct_j-\ct_k\right)}{4 f_j^2 (f_j^2-f_i^2) \left(f_j-f_k\right)} \Bigg] \,.
\end{split}
\end{align}
At first sight, denominators such as $(f_i\pm f_j)$ could seem problematic when the same or opposite eigenvalues are taken into account. However, when carefully considering the limit $f_j \rightarrow \pm f_i$ of $R_l^{\rm sym}$, it is possible to see that these denominators give rise to no divergences at all; for instance, some terms clearly lead to the derivative of $H$ in such a limit.
Similarly, when considering the weak-field expansion of these expressions, as is done in App.~\ref{app:weak_field_final}, no such denominator remains.

The full result present in Eq.~\eqref{eq:result} is lengthy and, therefore, not straightforward to interpret. A natural way to simplify its  structure is to assume further  properties for the field. For example, one could consider a background for which the different components possess the same spacetime dependence. Alternatively, one could consider parallel electric and magnetic fields, or fields of equal amplitude, for which the eigenvalues of the field strength greatly simplify, as can be seen from the corresponding expressions in App.~\eqref{eq:eigen} for the four-dimensional case.
Notable cases where these assumptions apply are the totally electric or magnetic field with a single polarisation, for which we can respectively  write $(\vec{E}=E(x)\,\vec{e}_1,\, \vec{B}=0)$, and $(\vec{E}=0,\, \vec{B}=B(x)\,\vec{e}_1)$. 
These configurations have been previously considered in the literature, cf. Refs.~\cite{Lee:1989vh, Gusynin:1998bt}, as well as the particular inhomogeneous, yet solvable background presented in Ref.~\cite{Dunne:2004nc} for fermionic QED.
We show the simplification that occurs in our setup under these assumptions in App.~\ref{app:phiF}, and find agreement with the cited literature.


\subsection{Derivative corrections to the Schwinger effect}\label{sec:Schwinger}
A particularly influential application of our results is the Schwinger effect, which was already briefly mentioned when discussing the renormalisation process in Sec.~\ref{sec:n0} and we will illustrate in the following. An information that we can extract from the full result is the analytic structure of the integrand in the propertime, in particular the position of the poles, if any. As mentioned earlier, the necessity to circumvent them in order to obtain a well-defined result implies that the effective action will develop an imaginary part. This contribution can be interpreted as a vacuum instability, what can be seen from the transition amplitude
\begin{align}
 \langle 0_{\rm out }\vert 0_{\rm in}\rangle   =e^{\mathi S_{\rm eff}}
\end{align}
between $\vert 0_{\rm in}\rangle$ and $\vert 0_{\rm out} \rangle$, i.e. between the in and out vacua.  
The imaginary part of the effective action generates an exponential decay in the corresponding probability, 
so that one can define the particle pair creation probability $P$ as 
\begin{align}
 1-P\equiv e^{-2\operatorname{Im} S_{\rm eff}}.
\end{align}
Importantly, note that the Schwinger effect is entirely of non-perturbative nature, given that it is lost once a weak-field expansion is performed.

Taking into account the $n=0$ contribution discussed in Sec.~\ref{sec:n0}, our  first contribution to the Schwinger effect from a purely electric field in $d=4$ reads
\begin{align}\label{eq:Schwinger}
\mathcal{L}_{\rm HE, ren} \overset{d=4}{=}& 
    \frac{E}{(4\pi)^2}
    \int_0^\infty \frac{d\tau}{\tau^2} 
    \left( \frac{1}{\sin(E\tau)} -\frac{1}{E \tau} -\frac{E \tau}{6} \right) e^{-M^2\tau},
\end{align}
which has poles at $\tau_j=\frac{j\pi}{E}$, $j\in\mathbb{N}-\{0\}$. They give rise to contributions to the production rate which are proportional to $\exp\left(-  j \pi\frac{M^2}{E}\right)$, resulting in a transseries in the variable $E$, i.e. series which generalize the notion of power series. For the interested reader, Ref.~\cite{Dunne:2014bca} provides the essential concepts on transseries and several relevant references.
In any case, the obtained production rate corresponds exactly to the result obtained for a constant electric field in the Heisenberg--Euler approximation~\cite{Dunne:2004nc}, as could have been guessed.

Considering now the derivative corrections that we have found, the exponential with the mass is once more present and guarantees the infrared convergence of the integral. Interestingly, poles arise just from the determinant prefactor and the $\coth$ functions in the $R_i$ factors, and are found to appear at the same values $\tau_n$ as in Eq.~\eqref{eq:Schwinger}.
The additional contributions to  the pair production rate for weakly varying fields correspond thus to a modification of the prefactor of the exponentials, i.e. it does not require an adjustement of the transseries expansion.


\section{Conclusion}\label{sec:conclusions}
In this work we have extended the applicability of the Covariant Derivative Expansion (CDE)---already known for its advantages across a wide class of models---to obtain resummed expressions for a quantum field theory in the presence of strong background fields. It builds upon the Gaillard--Cheyette trick, which  allows for a gauge-invariant separation of the background contributions and a grading in the number of derivatives acting on the field strength. This strong-field CDE method has been applied to derive the effective action for an electromagnetic field when a scalar field is integrated out, pointing out the strong-field and non-perturbative character of the expansion, as shown in Eq.~\eqref{eq:expansion}.

In the case of scalar QED, we have analysed in detail the first terms in a derivative-expansion of the field strength.
From the zeroth order contribution, we recover  the expected Heisenberg--Euler effective action, viz. Eq.~\eqref{eq:n0}; in  usual perturbative expansions, i.e. in the large-mass computations, this zeroth order corresponds to a resummation of the contributions involving only powers of the field strength (i.e., without derivatives).  The first corrections to the Heisenberg--Euler result requires the resummation of terms having an arbitrary number of field strengths and two derivatives acting on them. Even if the final result is lengthy, we have limpidly obtained closed expressions for these corrections, which are condensed in Eq.~\eqref{eq:result}. Importantly, some of our discussions, in particular Eqs.~\eqref{eq:n1} and \eqref{eq:n2}, describe a method that can be actually applied to  more general scenarios, for example with the inclusion of a  scalar background.

Taking into account the complexity of the computations performed, which involve a number of terms of order $10^3-10^4$, and the consequently increasing risk of errors, we have thoroughly checked our results with the sources available in the literature. In particular, our weak-field expansion can be seen to agree with those computed using a perturbative worldline approach~\cite{Fliegner:1997rk}, as well as with those coming from  perturbative ~\cite{vandeVen:1997pf,Franchino-Vinas:2024wof} and non-perturbative \cite{Franchino-Vinas:2023wea} heat kernel computations; in contrast, discrepancies are present when compared with the non-perturbative worldline computation in Ref.~\cite{Gusynin:1998bt}. Moreover, specific fields for which the result greatly simplifies have also been considered, viz. App.~\ref{app:phiF} for a field with a fixed polarisation. Curiously, this result agrees with the formulae in Ref.~\cite{Gusynin:1998bt}, suggesting that the discrepancies observed for weak fields do not affect this type of background.

An immediate application of our findings pertains the Schwinger effect.
Using the second-derivative contributions to the effective Lagrangian, one can predict corrections to the probability of pair creation. We have shown that, at this order, the analytic structure of the expansion is equivalent to that in the Heisenberg--Euler expansion, implying in its turn that the involved transseries is of the same type, with modifications taking place at the level of the corresponding coefficients. A notable enhancement of the Schwinger effect is thus not present at this order.

Even if we have focused on the computation of the effective action, in principle one could employ the strong-field CDE to obtain expansions of two-point functions. This would introduce a further scale to the theory, namely the distance between the arguments of the Green function, which will have to be appropriately taken into account  in our expansions; this would allow us to study the stress-energy tensor, as done in Refs.~\cite{Boasso:2025ofd, Boasso:2025ptx} for a gravitational scenario. Preliminary calculations in this direction are promising.

Natural extensions of this work include the consideration of quantum fields with spin. A specific potential $U$ can also be considered, either as a perturbation like in Sec. \ref{sect:expansion}, or including the first terms of $\tilde U$ appearing in Eq.~\eqref{eq:U_tilde} into the propagator. We expect that a closed expression could be readily obtained up to second-order in derivatives and the main results for Yukawa interactions studied in \cite{Franchino-Vinas:2023wea} could be rederived. Furthermore, it would be intriguing to see if the method studied in this article is applicable to perform a resummation in the mixed Yukawa and electromagnetic scenario, which is missing so far.
On the other side, the introduction of a curved spacetime background is currently being analysed adapting the considerations in Ref.~\cite{Larue:2023uyv} and will be left for future work. This would offer a further point of view in recent lively discussions regarding the nature of pair creation in gravitational backgrounds~\cite{Wondrak:2023zdi, Ferreiro:2023jfs, Akhmedov:2024axn, Boasso:2024ryt, Garani:2025qnm, vanSuijlekom:2025tjt}. 

Of course, one could wonder if other terms can be resummed in the effective action. At first sight, this seems a non-trivial task in this approach, because it will involve solving a differential equation containing a source with  cubic or higher powers of the variable. However, certain cases could be tractable if further assumptions on the field are considered. In particular, it would be interesting to analyse if an enhancement on observables, such as the Schwinger effect, could be present. Indeed, fast varying  field strengths---not necessarily in a strong-field context---can also induce the creation of pairs~\cite{Brezin:1970xf, Boasso:2024ryt, VicenteGarcia-Consuegra:2025lkh}. Rapidly varying fields can also be employed as catalysers of pair creation in the presence of an intense background~\cite{Schutzhold:2008pz}. For a given field configuration, by taking some particular scalings on it and its derivatives, one could also study the combined effect of fast variations and large fields and investigate  whether a self-assisted mechanism might take place.  

Lastly, from the computational point of view it would be helpful to derive, if possible, the Hilbert series corresponding to our two scales EFT approach, namely without imposing onshellness and at the resummed level of the basic operators. The availability of the Hilbert series would already provide a first intuition on the level of complexity of the computations needed at each level of the expansion.


\section*{Acknowledgments}
The authors acknowledge support from the Enigmass+ research federation (CNRS, Université Grenoble Alpes, Université Savoie Mont-Blanc).  This work is supported by the French National Research Agency in the framework of the “France 2030” (ANR-15-IDEX-02). The work of D.S. and J.Q. is supported by the EFFORT project, funded through the IRGA program of Université Grenoble Alpes (UGA), and by the Tremplin project from CNRS Physique.
SAF thanks the members of the LAPTh, where this project was conceived, and the Institut Denis Poisson, especially M. Chernodub, for their warm hospitality.  The research activities of SAF have been carried out in the framework of the INFN Research Project QGSKY, Project PIP 11220200101426CO from Consejo Nacional de Investigaciones Científicas y Técnicas (CONICET) and Project 11/X748 of UNLP. The authors would like to acknowledge the contribution of the COST Action CA23130.
The authors also extend their appreciation to the Italian National Group of Mathematical Physics (GNFM, INdAM) for its support.  The authors lastly acknowledge fruitful discussions and funding from the workshops ``New Trends in First Quantisation: Field Theory, Gravity and Quantum Computing'' (Heraeus Stiftung) and ``First Quantisation for Physics in Strong Fields'' (University of Edinburgh).


\appendix
\section{Weak-field expansion}
\subsection{Minimal basis}\label{sec:minimal_basis}
Any scalar operator of a given mass dimension built from contractions of the field strength and its derivatives can be decomposed in a minimal basis using the symmetries in the indices, Bianchi identities and integration by parts. Here we give a minimal basis for operators of mass dimension 6 (only one element, denoted $o_3$) and of mass dimension 10 containing 2 partial derivatives (7 elements, $o_{5,i}$):
\begin{align}
    \begin{split}
        o_3&=\partial_\al F_{\mu\nu}\partial^\al F^{\mu\nu}\, ,
    \end{split}
    \begin{split}\label{eq:o5i}
        o_{5,1}&=F^{\al\be}F^{\mu\nu}\partial_\al F_{\nu\rho}\partial_\mu F_\be{}^\rho\, ,\\
        o_{5,2}&=F^{\be\al}F_\be{}^\sig \partial_\sig F_{\nu\rho}\partial_\al F^{\nu\rho}\, ,\\
        o_{5,3}&=F_\al{}^\mu F^{\al\be}F^{\nu\rho}\partial_\mu\partial_\be F_{\nu\rho}\, ,\\
        o_{5,4}&=F^{\al\be}F^{\mu\nu}\partial_\mu F_\al{}^\rho \partial_\nu F_{\be\rho}\, ,\\
        o_{5,5}&=F^{\al\be}F^{\mu\nu}\partial_\rho F_{\al\be}\partial^\rho F_{\mu\nu}\, ,\\
        o_{5,6}&=F_\al{}^\mu F^{\al\be}\partial_\rho F_{\mu\nu}\partial^\rho F_\be{}^\nu\, ,\\
        o_{5,7}&=F^{\al\be}F_{\al\be}\partial_\rho F_{\mu\nu}\partial^\rho F^{\mu\nu} \, .\\
    \end{split}
\end{align}
 We obtained them using the algorithm described in Ref.~\cite{Muller:1996cq}---this reference was also followed to reduce general operators to our bases.

\subsection{Weak-field expansion of the effective Lagrangian}\label{app:weak_field_Lagrangian}
The effective Lagrangian obtained when integrating out a scalar coupled to an electromagnetic background is well-known in its weak-field approximation. It corresponds to the expansion
\begin{align}
\begin{split}
\mathcal{L}_{\textrm{eff}}
&=\frac{\colred{1}}{(4\pi)^{d/2}}\int_0^\infty \frac{\dd \tau}{\tau^{1+d/2}}e^{-M^2\tau}\sum_{n=0}^\infty \tau^nO_n\\
&=\frac{\colred{1}}{(4\pi)^{d/2}}\sum_{n=0}^\infty \Gamma\left(n-\frac{d}{2}\right)\,M^{d-2n}O_n\,,
\end{split}
\end{align}
where $\Gamma(\cdot)$ is the gamma function and $O_n$ are the heat kernel---or Schwinger--DeWitt---coefficients of the differential operator inside the $\operatorname{Tr}\operatorname{Log}$  in Eq.\eqref{eq:TrLog}. These coefficients are made of invariants of the theory~\cite{Vassilevich:2003xt}, which in our case correspond to Lorentz scalars made of contractions of field strengths and derivatives thereof; see Ref.~\cite{Franchino-Vinas:2023wea} for more details. 

Here we are interested in listing the contributions to the first Schwinger--DeWitt coefficients in which two derivatives act on field strengths; the result is going to be expressed in the elements of the  basis introduced in App.~\ref{sec:minimal_basis}. The computation has been performed by taking the weak-field expansion of our strong-field result at different stages of the computations presented in this article, giving
\begin{align}\label{eq:O3}
    O_3 &= \frac{1}{120}o_3\, ,
\\
\label{eq:O5}
    O_5 &\supset \frac{1}{756} o_{5,1} - \frac{1}{1890}o_{5,2} - \frac{1}{1080}o_{5,3} + \frac{1}{630}o_{5,4} - \frac{1}{432} o_{5,5} + \frac{1}{630}o_{5,6} + \frac{1}{1440}o_{5,7} \, .
\end{align}
 One of the checks involved employing the usual CDE expansion, where the expansion utilises the free propagator as in Eq.~\eqref{eq:CDE-wf}
An alternative option is to expand our final result, as done in App.~\ref{app:weak_field_final}.
Note that there are further terms in $O_5$, e.g. $\partial^3 F\partial^3 F$, which goes beyond the order in the derivative expansion considered in Sec. \ref{sect:inhom}. 
In addition, recall that the Eqs.~\eqref{eq:O3} and \eqref{eq:O5} are valid up to integration by parts and Bianchi identities.

The Schwinger--DeWitt coefficients can also be found in Ref.~\cite{Fliegner:1997rk}, where they were computed through worldline techniques, and the result is summarised in Ref.~\cite{Navarro-Salas:2020oew}, where a ``non-minimal basis'' was employed. All of these results agree with ours once they are written in the same basis.
On the other hand, the result in Ref.~\cite{Gusynin:1998bt} does not match ours; a similar mismatch  was already noted in Ref.~\cite{Franchino-Vinas:2023wea}.


\subsection{Weak-field expansion of our final result}\label{app:weak_field_final}
We consider the weak-field expansions of the form factors in Eq.~\eqref{eq:result},
\begin{align}
\begin{split}
    R^{\text{sym}}_0(f_i,f_j)&=-\frac{\tau^3}{30}(f_i-f_j)+\frac{\tau^5}{420}(f_i-f_j)(f_i^2+f_j^2)+\mathcal{O}(\tau^7)\, ,
    \\
    9 R^{\text{sym}}_1(f_i,f_j,f_k)&= -\frac{\tau^3}{20}+\frac{\tau^5}{420}
    \left(2 f_i^2+2f_j^2+2f_k^3-51 f_i f_j+9f_k(f_i-f_j) \right) +\mathcal{O}(\tau^7)\, ,
    \\
    9 R^{\text{sym}}_2(f_i,f_j,f_k)&=  -\frac{2\tau^3}{5}+\frac{\tau^5}{840}
    \left(23 f_i^2+23f_j^2+20f_k^3-6 f_i f_j-12f_k(f_i-f_j) \right) +\mathcal{O}(\tau^7)\, .
\end{split}
\end{align}
It is now possible to reverse the diagonalisation by replacing $f_l^n$ by $(F^n)^{\lambda\chi}$, where the indices $\lambda$ and $\chi$ of a given $l$ are those in the corresponding $A_{(l)}$ in Eq.~\eqref{eq:result}.
The expressions in App.~\ref{app:weak_field_Lagrangian} are easily obtained from these limits, taking into account the contribution from the determinant.

We can readily compare these expansions with the results contained in Ref.~\cite{Gusynin:1998bt}, after reducing them to our choice of tensor contractions. We agree on the form factor $R_0^{\rm sym}$, while for $R_1^{\rm sym}$ and $R_2^{\rm sym}$ we observe a disagreement, which could be the origin of our weak-field discrepancy. For the sake of completeness, we include the expansions that we obtain using the findings in Ref.\cite{Gusynin:1998bt}:
\begin{align}
\begin{split}
    9 R^{\rm Gus}_1(f_i,f_j,f_k)&= -\frac{\tau^3}{30}+\frac{\tau^5}{630}
    \left(2 f_i^2+2f_j^2+2f_k^3-153 f_i f_j+9f_kf_i-51f_kf_j) \right) +\mathcal{O}(\tau^7)\, ,
    \\
    9 R_2^{\rm Gus}(f_i,f_j,f_k)&=  -\frac{2\tau^3}{5}+\frac{\tau^5}{630}
    \left(20 f_i^2+20f_j^2+17f_k^3-9 f_i f_j-9f_k(f_i-f_j) \right) +\mathcal{O}(\tau^7) \, .
\end{split}
\end{align}


\section{Definitions for the diagonalisation}\label{app:diagoF}
We recall here a few definitions, taken from Ref.~\cite{Gusynin:1998bt}, which are relevant to the diagonalisation of a general field strength $F$ in four dimensions:
\begin{align}
    {\,}^{*}\!F_{\mu\nu}&\equiv\frac{1}{2}\eps_{\mu\nu\rho\sig}F^{\rho\sig}\, ,
\\
  \F&\equiv-\frac{1}{4}F^{\mu\nu}F_{\mu\nu} \,,
    \\
\G&\equiv\frac{1}{8}\eps^{\mu\nu\rho\sig}F_{\mu\nu}F_{\rho\sig}\, ,
\\
\label{eq:eigen}
    K_\pm&\equiv\sqrt{\sqrt{\F^2+\G^2}\pm\F}\, ,
\\
    f_j&\equiv(\mathi K_-, -\mathi K_-, K_+ , - K_+)_j\, ,\\
    \fb_j&\equiv(- K_+,K_+,-\mathi K_- , \mathi K_-)_j\, ,
\\
A_{(j)\mu\nu}&=\frac{-\fb^2_j \eta_{\mu\nu}+f_j F_{\mu\nu}+F^2_{\mu\nu}-\mathi \fb_j {\,}^{*}\!F_{\mu\nu}}
    {2(f_j^2-\fb^2_j)}
    \,.
\end{align}
${\,}^{*}\!F_{\mu\nu}$ is the Hodge dual of $F_{\mu\nu}$, while $\F$ and $\G$ are the fundamental scalars that are quadratic in the field strength; on the other hand, $K_{\pm}$, $f_j$, $\fb$ and $A_{(j)}$ are related to the diagonalisation procedure introduced in Eq.~\eqref{eq:F_diagonalisation}. It should be clear that setting $\G$ or $\F$ to zero simplifies the expression of the eigenvalues $f_j$.


\section{Example with a simplified field strength} \label{app:phiF}
In Refs.~\cite{Lee:1989vh, Gusynin:1998bt}, a particular case is considered where the field strength reads
\begin{align}\label{eq:phiF}
    F^{\mu\nu}(x) = \Phi(x) \mathds{F}^{\mu\nu} \,.
\end{align}
This corresponds to having electric and magnetic fields with fixed polarisations and a constant ratio of intensities.
In this appendix, we will show how our general result simplifies by assuming Eq.~\eqref{eq:phiF}.
The additional constraint $\G=0$, which is particularly useful when considering a vanishing electric or magnetic field, has also been used to get a compact result in Refs.~\cite{Lee:1989vh, Gusynin:1998bt}; we will make use of it only when explicitly said.


\subsection{Tensor simplification}\label{app:tensor_simpli}
We will see now the simplification of the tensorial structures in Eq.~\eqref{eq:result} once the field strength in Eq.~\eqref{eq:phiF} is assumed.
First of all, a useful equation under this hypothesis is
\begin{align}
    \Big(-2\F \,\eta^{\mu\nu}+\left(F^2\right)^{\mu\nu}\Big)\partial_\nu\Phi=0 \,,
\end{align}
which is obtained from the Bianchi identity for the field strength.
The tensorial structures can thus be recast as
\begin{align}
\begin{split}\label{eq:AA4phi}
    A_{(i)}^{\mu\al}&A_{(j)}^{\nu\be}\partial_\al\partial_\be F_{\mu\nu}
    =\frac{1}{8 \Phi  \left(f_i^2-\fb_i^2\right) \left(f_j^2-\fb_j^2\right)}
    \bigg[4\partial^\mu{}_\mu\Phi  \mathcal{F} \Big(f_j \left(2 \mathcal{F}-\fb_i^2\right)+f_i \left(\fb_j^2-2 \mathcal{F}\right)\Big)
    \\
    &+\mathi \eps_{\mu\nu\rho\sig} \partial^{\mu\al}\Phi F_\al{}^\nu F^{\rho\sig} \left(\fb_i-\fb_j\right) \left(\fb_i \fb_j+2
   \mathcal{F}\right)
   \bigg]\, ,
\end{split}
\end{align}
\begin{align}
\begin{split}
    A_{(i)}^{\al\mu}&A_{(j)}^{\be\rho}A_{(k)}^{\nu\sig}\partial_\al F_{\mu\nu}\partial_\be F_{\rho\sig}
    =\frac{f_i f_j}{16 \Phi ^2 \left(f_i^2-\bar{f}_i^2\right)
   \left(f_j^2-\bar{f}_j^2\right) \left(f_k^2-\bar{f}_k^2\right)}\\
   \times&\Bigg\lbrace2 \partial^\mu\Phi\partial_\mu\Phi \bigg[2 \mathcal{F} \Big(f_j f_k \left(\bar{f}_i^2-2 \mathcal{F}\right)+ f_i f_k \left(2 \mathcal{F}-\bar{f}_j^2\right)+ f_i f_j \left(\bar{f}_k^2-2 \mathcal{F}\right)\Big)\\
   &+\bar{f}_i \bar{f}_j \Big(-8
   \mathcal{F}^2+ \sum_l f_l^4 \Big)-\left(2 \mathcal{F}+\bar{f}_i^2\right) \left(2 \mathcal{F}-\bar{f}_j^2\right) \left(2 \mathcal{F}-\bar{f}_k^2\right)\bigg]\\
   &+\mathi \eps_{\mu\nu\rho\sig} \partial^{\mu}\Phi \partial^{\al}\Phi F_\al{}^\nu F^{\rho\sig} \Big(-f_k \left(\bar{f}_i-\bar{f}_j\right) \left(2
   \mathcal{F} +\bar{f}_i \bar{f}_j+\right)+f_j \left(\bar{f}_i+\bar{f}_k\right) \left(2 \mathcal{F}-\bar{f}_i \bar{f}_k\right)\\
   &\hspace{8.5cm} +f_i \left(\bar{f}_j-\bar{f}_k\Big) \left(2 \mathcal{F} + \bar{f}_j \bar{f}_k\right)\right)\Bigg\rbrace\, ,
\end{split}
\end{align}
\begin{align}
\begin{split}
    A_{(i)}^{\al\rho}&A_{(j)}^{\be\mu}A_{(k)}^{\nu\sig}\partial_\al F_{\mu\nu}\partial_\be F_{\rho\sig}
    =\frac{f_i f_j}{16 \Phi ^2 \left(f_i^2-\fb_i^2\right) \left(f_j^2-\fb_j^2\right) \left(f_k^2-\fb_k^2\right)}\\
    \times&\Bigg\lbrace-2 \partial^\mu\Phi\partial_\mu\Phi \Big[2 \F\bigg( f_j f_k \left(\fb_i^2-2 \F\right)+ f_i f_k \left(2 \F-\fb_j^2\right)+f_i f_j \left(2 \F-\fb_k^2\right)\Big)+8 \F^3\\
    &+\fb_i \fb_j\Big(8
   \F^2-\sum_l f_l^4\Big)+\fb_k^2 \left(2 \F-\fb_i^2\right) \left(2 \F-\fb_j^2\right)-2 \F \fb_i^2 \fb_j^2+4 \F^2 \fb_i^2+4 \F^2 \fb_j^2\bigg]\\
   &+\mathi \eps_{\mu\nu\rho\sig} \partial^{\mu}\Phi \partial^{\al}\Phi F_\al{}^\nu F^{\rho\sig} \Big(f_k \left(\fb_i-\fb_j\right) \left(\fb_i \fb_j+2 \F\right)+f_j \left(\fb_i-\fb_k\right) \left(\fb_i \fb_k+2 \F\right)\\
   &\hspace{8.5cm}+f_i \left(\fb_j+\fb_k\right) \left(2 \F-\fb_j
   \fb_k\right)\Big)\Bigg\rbrace\, .
\end{split}
\end{align}
After integration by parts in Eq.~\eqref{eq:AA4phi}, we see that the same two tensor structures are present in the three equations: $\partial^\mu\Phi\partial_\mu\Phi$ and $\eps_{\mu\nu\rho\sig} \partial^{\mu}\Phi \partial^{\al}\Phi F_\al{}^\nu F^{\rho\sig}$.
The latter possesses the peculiarity that it vanishes when taking $\G=0$.
Indeed, any 2-form can be written as a sum of two simple parts, and if $\epsilon^{\mu\nu\rho\sig}F_{\mu\nu}F_{\rho\sig}=0$, then $F$ must be simple, i.e.
\begin{align}
    F^{\mu\nu}\Big|_{\G=0} = a^{[\mu}b^{\nu]} \,,
\end{align}
where $a$ and $b$ are two appropriate vectors (see \cite[p. 93]{Misner:1973prb}).
From this decomposition, we trivially obtain
\begin{align}
    \eps_{\mu\nu\rho\sig}F_\al{}^\nu F^{\rho\sig}\Big|_{\G=0}=0 \,.
\end{align}


\subsection{Eigenvalue simplification}
Replacing the expressions  obtained in App.~\ref{app:tensor_simpli} into the effective action \eqref{eq:result}, we can evaluate the sums over the eigenvalues. As mentioned earlier,  a direct replacement of the eigenvalues give rise to ill-defined expressions; we rather need to take the corresponding limits. We will limit ourselves to one situation that greatly simplify the expressions.

Choosing $\G=0$, we normalise $\mathds{F}^{\mu\nu}\mathds{F}_{\mu\nu}=2$, such that $\Phi=\sqrt{-2\F}$, and compute
\begin{align}
\begin{split}\label{eq:sumR0}
    &\sum_{i,j}R_0^{\rm sym}(f_i,f_j)
    A_{(i)}^{\mu\al}A_{(j)}^{\nu\be}\partial_\al\partial_\be F_{\mu\nu}\Big|_{\G=0} =\frac{\partial^\al{}_\al\Phi}{4\Phi^2}
    \left(\Phi\tau-\frac{3}{\tan(\Phi\tau)} +\frac{3\Phi\tau}{\tan^2(\Phi\tau)} \right)\, ,
\end{split}
\\[20pt]
\begin{split}
    &\sum_{i,j,k}R_1^{\rm sym}(f_i,f_j,f_k)
    A_{(i)}^{\al\rho}A_{(j)}^{\be\mu}A_{(k)}^{\nu\sig}\partial_\al F_{\mu\nu}\partial_\be F_{\rho\sig}\Big|_{\G=0}\\
    &\hspace{2cm}= \frac{\partial_\al\Phi\partial^\al\Phi}{4 \Phi^3}\left(-\Phi\tau+\frac{-2+3(\Phi\tau)^2}{\tan(\Phi\tau)}
    -\frac{2\Phi\tau}{\tan^2(\Phi\tau)}+\frac{4(\Phi\tau)^2}{\tan^3(\Phi\tau)} \right) \,,
\end{split}
\\[20pt]
\begin{split}
    &\sum_{i,j,k}R_2^{\rm sym}(f_i,f_j,f_k)
    A_{(i)}^{\al\rho}A_{(j)}^{\be\mu}A_{(k)}^{\nu\sig}\partial_\al F_{\mu\nu}\partial_\be F_{\rho\sig}\Big|_{\G=0}\\
    &\hspace{2cm}=\frac{\partial_\al\Phi\partial^\al\Phi}{4 \Phi^3}
    \left(-\Phi\tau+\frac{-1+2(\Phi\tau)^2}{\tan(\Phi\tau)}
    -\frac{\Phi\tau}{\tan^2(\Phi\tau)}+\frac{2(\Phi\tau)^2}{\tan^3(\Phi\tau)} \right)\,.
\end{split}
\end{align}
The determinant contribution in Eq.~\eqref{eq:result}  simply becomes $\frac{\Phi \tau}{\sinh(\Phi \tau)}$, and we can proceed with the integration by parts of Eq.~\eqref{eq:sumR0}. Assembling all the pieces together, for the field in Eq.~\eqref{eq:phiF} and $\G=0$, we obtain
\begin{align}
\begin{split}
    \mathcal{L}_{\mathrm{eff}}^{\partial^2}\Big|_{\G=0}&=
    \frac{\colred{-1}}{(4\pi)^{2}}\int_0^\infty \frac{\dd\tau}{\tau} e^{-M^2\tau}
    \frac{\partial_\al\Phi\partial^\al\Phi}{4\Phi \sin(\Phi\tau)} \left(1-\frac{2\Phi\tau}{\tan(\Phi\tau)}
    +\frac{3}{\tan^2(\Phi\tau)}+\frac{3\Phi\tau}{\tan^3(\Phi\tau)} \right)\,.
\end{split}
\end{align}
Notice that in the presence of only an electric field, $\Phi=E$, while for a magnetic field, $\Phi=\mathi B$. This result agrees with the expression provided in Ref.~\cite{Gusynin:1998bt}.

\bibliographystyle{JHEP}
\bibliography{biblio}
\end{document}